\documentclass[traditabstract]{aa}
\usepackage{txfonts}
\usepackage{graphicx}
\usepackage{natbib}
\usepackage{color}
\usepackage{longtable}

\begin{document}

\title{Thermal infrared properties of classical and type II Cepheids}
\titlerunning{Thermal infrared properties of Cepheids}

\subtitle{Diffraction limited 10\,$\mu$m imaging with VLT/VISIR\thanks{Based on observations made with ESO telescopes at Paranal observatory under program ID 081.D-0165(A)}}

\author{ A.~Gallenne\inst{1} \and  
  P.~Kervella\inst{1} \and
  A.~M\'erand\inst{2} }
  
\authorrunning{A. Gallenne et al.}

\institute{LESIA, Observatoire de Paris, CNRS UMR 8109, UPMC, Universit\'e
  Paris Diderot, 5 Place Jules Janssen, F-92195 Meudon, France
  \and European Southern Observatory, Alonso de C\'ordova 3107,
  Casilla 19001, Santiago 19, Chile}
  
  \offprints{A. Gallenne} \mail{alexandre.gallenne@obspm.fr}
       
   \date{Received May 20, 2011; accepted November 24, 2011}
       
\abstract
	{We present new thermal IR photometry and spectral energy distributions (SEDs) of eight classical Cepheids (type~I) and three type~II Cepheids, using \emph{VISIR} thermal IR photometric measurements, supplemented with literature data. We used the BURST mode of the instrument to get diffraction-limited images at 8.59, 11.25 and 11.85\,$\mu$m. The SEDs show a IR excess at wavelengths longer than $10\mu\mathrm{m}$ in ten of the eleven stars. We tentatively attribute these excesses to circumstellar emission created by mass loss from the Cepheids. With some hypotheses for the dust composition, we estimated a total mass of the envelope ranging from $10^{-10}$ to $10^{-8}\,M_\odot$. We also detect a spatially extended emission around AX~Cir, X~Sgr, W~Sgr, Y~Oph and U~Car while we do not resolve the circumstellar envelope (CSE) for the other stars. The averaged circumstellar envelope brightnesses relative to the stellar photosphere are $\alpha (\mathrm{AX~Cir}) = 13.8 \pm 2.5\,\%, \alpha (\mathrm{X~Sgr}) = 7.9 \pm 1.4\,\%, \alpha (\mathrm{W~Sgr}) = 3.8 \pm 0.6\,\%, \alpha (\mathrm{Y~Oph}) = 15.1 \pm 1.4\,\%$ and $\alpha (\mathrm{U~Car}) = 16.3 \pm 1.4\,\%$ at 8.59\,$\mu$m. With this study, we extend the number of classical Cepheids with detected CSEs from 9 to 14, confirming that at least a large fraction of all Cepheids are experiencing significant mass loss. The presence of these CSEs may also impact the future use of Cepheids as standard candles at near and thermal infrared wavelengths.}


\maketitle


\section{Introduction}

\citet{Kervella-2006-03} discovered in the $N$ and $K$ band a circumstellar envelope (CSE) around $\ell$~Car using the MIDI and VINCI instruments from the VLTI. Its typical size is $2-3\,R_\star$ with a contribution of 4\,\% to the total flux in $K$. Similar interferometric detections were then reported for other Cepheids \citep{Merand-2006-07, Merand-2007-08} leading to the conclusion that a significant fraction of all Cepheids are surrounded by a CSE. From mid- and far-IR observations with the Spitzer telescope, \citet{Barmby-2010-11} also detected extended emission around a significant fraction of their sample of 29 classical Cepheids. The case of $\delta$~Cep and its extended emission is well discussed in \citet{Marengo-2010-12}. These CSEs have an effect on the infrared surface brightness technique (IRSB) since the Cepheid appears brighter and also on interferometric measurements since the star appears larger than it really is. It is therefore necessary to quantify this excess (linked to an extended emission) in order to estimate the bias on the distance. This is particularly important for the James Webb Space Telescope (JWST) that will be able to observe Cepheids in distant galaxies. Their distances could be estimated via the use of IR P--L relations but the presence of infrared excess will degrade the distance accuracy.

This circumstellar material is also important in the context of Cepheid mass-loss since it may play a significant role in the problem of the Cepheid mass discrepancy \citep[][...]{Neilson-2010-11}. The infrared excess could be linked to a mass-loss mechanism generated by shocks between different layers of the Cepheid's atmosphere during the pulsation cycle. A correlation between the period and the envelope flux (relatively to the star) was proposed by \citet{Merand-2007-08}. A Cepheid with a larger pulsation period would have a larger IR excess. On the other hand, from photometry on a larger sample of Cepheids, \citet{Neilson-2010-06} reach a different conclusion. It is thus essential to study these CSEs to quantify their contribution and to understand how they form.

To progress on this question, we present new observations of a few classical (type I) and type II Cepheids with the \emph{VISIR} instrument of the VLT. The mid-IR wavelength coverage of this instrument is well suited to study the infrared contribution of the CSEs. 
The instrument configurations and the data reduction methods are detailed in Sect.~\ref{observation_and_data_processing}. In Sect.~\ref{data_analysis} we present the data analysis using aperture photometry applied to our images and we study the spectral energy distribution of our sample of stars. We also look for a spatially resolved component using a Fourier technique analysis. Finally we discuss our results in Sect.~\ref{discussion}.


\section{Observations and data processing}
\label{observation_and_data_processing}

The selected sample of classical Cepheids was chosen for their brightness, the range of pulsation period (from 4 to 45 days) and their angular size so that they are resolvable by long-baseline interferometry. Y~Oph seems to exhibit the brightest CSE detected around a Cepheid \citep{Merand-2007-08} and the goal is to explore larger distances from the Cepheid. The type II Cepheids  we selected have been extensively studied and are known to have strong IR excesses linked to their CSEs. They will be used in our analysis both to compare the properties of their CSEs with those of classical Cepheids, and to validate our modeling approach to estimate the thermal infrared excess (based on the results of previous works on type II Cepheids). We present in Table~\ref{cepheid_parameter} some relevant parameters for our sample of observed Cepheids.

The observations were performed using the ESO mid-infrared instrument \emph{VISIR} installed on UT3 at the VLT (Paranal, Chile). This instrument works in the $N$ ($8-13\,\mu$m) and $Q$ ($16-24\,\mu$m) atmospheric windows and provides imaging and long-slit spectroscopy. As we need the best spatial resolution, we chose the BURST mode \citep{Doucet-2006-} of \emph{VISIR} to overcome the atmospheric seeing. In this mode, short exposure images (i.e. faster than the atmospheric coherence time) are taken in order to freeze the turbulence. Each short exposure frames so presents one principal speckle. With sets of thousands of short exposure frames we can apply a shift-and-add process. This enhances the quality of the data and enabled us to reach the diffraction limit of the telescope. To correct for instrumental artefacts and background thermal emission, we applied the classical chop-nod technique. 

The raw data of the \emph{VISIR} BURST mode are data cubes in which the star appears in a different position in the field (the chop-nod positions, perpendicular each other) for each frames. Our first step consisted of a classical subtraction of the chopped and nodded images in order to remove the thermal background and in storing them in data cubes that contain thousands of frames ($\sim 20\,000$). To have the best diffraction-limited images, we selected $50\,\%$ of the best frames according to the brightest pixel (as tracer of the Strehl ratio). We then proceeded to a precentering (at a integer pixel level), a spatial resampling by a factor of 4 using a cubic spline interpolation and a fine recentering using a Gaussian fitting (at a precision level of a few milliarcseconds). The resulting cubes were then averaged to get the final image used in the data analysis process (see Sect.~\ref{data_analysis}). This raw data processing has already been used and has proven its efficiency \citep[see e.g. ][]{Kervella-2007-11,Kervella-2009-05,Gallenne-2011-03}.

\begin{table*}[ht]
\centering
\caption{Some relevant parameters of our Cepheids.}
\begin{tabular}{cccccccc} 
\hline
\hline
Stars	  & $P$\tablefootmark{a}	& 	$\mathrm{MJD_0}$\tablefootmark{b}	& $<V>$\tablefootmark{c}& $<K>$\tablefootmark{d}	& $\theta$\tablefootmark{e}	& $\pi$\tablefootmark{f}	& Type \\
		  &	(days)						&						&						&			& 				(mas)				& 	(mas)		&					\\
\hline
FF~Aql				&  4.4709			& 	41575.928		&	5.37			&	3.49	&	$0.88\,\pm\,0.05$	&	$2.81\,\pm\,0.18$	& I	\\
AX~Cir				&  5.2733			& 	51646.100		&	5.88			&	3.76	&	$0.70\,\pm\,0.06$	&	$3.22\,\pm\,1.22$	& I	\\
X~Sgr				&  7.0128			& 	51653.060		&	4.55			&	2.56	&	$1.47\,\pm\,0.04$	&	$3.00\,\pm\,0.18$	& I	\\
$\eta$~Aql		&  7.1767			& 	36084.156		&	3.90			&	1.98	&	$1.84\,\pm\,0.03$	&	$2.78\,\pm\,0.91$	& I	\\
W~Sgr				&  7.5950			& 	51652.610		&	4.67			&	2.80	&	$1.31\,\pm\,0.03$	&	$2.28\,\pm\,0.20$	& I	\\
Y~Oph				&  17.1242		& 	51652.820		&	6.17			&	2.69	&	$1.44\,\pm\,0.04$	&	$2.04\,\pm\,0.08$	& I	\\
U~Car				&  38.8124		& 	51639.740		&	6.29			&	3.52	&	$0.94\,\pm\,0.06$	&	$2.01\,\pm\,0.40$	& I	\\
SV~Vul				&  45.0121		& 	43086.390		&	7.22			&	3.93	&	$0.80\,\pm\,0.05$	&	$0.79\,\pm\,0.74$	& I	\\
\hline
R~Sct				&  146.50			&	44871.500		&  6.70			&	2.27	&	--								&	$2.32\,\pm\,0.82$	& II	\\
AC~Her				&  75.010			&	35097.300		&	7.90			&	5.01	&	--								&	$0.70\,\pm\,1.09$	& II	\\
$\kappa$~Pav	&	9.0814			&	46683.569		&	4.35			&	2.79	&	$1.17\,\pm\,0.05$	&	$6.00\,\pm\,0.67$	& II	\\
\hline
\end{tabular}
\tablefoot{$P$ is the pulsation period and $\mathrm{MJD_0}$ denotes the reference epoch ($\mathrm{MJD_0}$ = $\mathrm{JD_0}$ - 2400000.5). $<V>$ and $<K>$ are the mean apparent $V$ and $K$ magnitudes, $\theta$ the angular diameter and $\pi$ the trigonometric parallax. Type denotes the classical (type I) or type II Cepheids.}
\begin{flushleft}
\tablefoottext{a}{from \citet{Feast-2008-06-2} for $\kappa$~Pav ; from \citet{Samus-2009-01} for the others.} \\
\tablefoottext{b}{from \citet{Samus-2009-01} for SV~Vul, FF~Aql, $\eta$~Aql, R~Sct and AC~Her ; from \citet{Feast-2008-06-2} for $\kappa$~Pav ; from \citet{Berdnikov-2001-12} for the others.} \\
\tablefoottext{c}{from \citet{Fernie-1995-01} for the classical Cepheids ; from \citet{Samus-2009-01} for the type II.} \\
\tablefoottext{d}{from \citet{Welch-1984-04} for FF~Aql and X~Sgr ; from \emph{DENIS} for AX~Cir and W~Sgr ; from \citet{Barnes-1997-06} for $\eta$~Aql ; from \citet{Laney-1992-04} for Y~Oph, U~Car and SV~Vul ; from \citet{Taranova-2010-02} for R~Sct and AC~Her ; from \citet{Feast-2008-06-2} for $\kappa$~Pav.} \\
\tablefoottext{e}{limb-darkened angular diameters from \citet{Kervella-2004-03a} for X~Sgr, $\eta$~Aql, W~Sgr and Y~Oph ; from \citet{Groenewegen-2007-11} for FF~Aql ; predicted diameter from \citet{Moskalik-2005-06} for AX~Cir ; from \citet{Feast-2008-06-2} for $\kappa$~Pav ; from \citet{Groenewegen-2008-09} for the others.} \\
\tablefoottext{f}{from \citet{Benedict-2007-04} for FF~Aql, X~Sgr and W~Sgr ; from \citet{Hoffleit-1991-} for U~Car ; from \citet{Merand-2007-08} for Y~Oph; from \citet{Perryman-1997-07} for the others.}
\end{flushleft}
\label{cepheid_parameter}
\end{table*}

The observations were carried out on the nights of 2008 May 23--24. Table~\ref{log} lists the sequence of our observations including the reference stars that were observed immediately before and after the Cepheids in the same instrumental setup. Series of observations were obtained in three filters: PAH1, PAH2 and SiC (respectively $8.59\,\pm\,0.42\,\mu$m, $11.25\,\pm\,0.59\,\mu$m and $11.85\,\pm\,2.34\,\mu$m). We chose the smallest pixel scale of 75\,mas/pixel (before resampling) to ensure a proper sampling of the FWHM of the Airy pattern (average FWHM $\sim$ 4 pixels for the reference stars). For an unknown reason, the nodding position was out of the detector for the observations \#15, \#17, \#19 to \#22, \#25 to \#28 and \#31 to \#34 but we still have enough frames ($\sim 10\,000$) to correctly subtract the background emission. The observations \#23 and \#24 were chopped and nodded out of the detector and were not included in the analysis. Due to the lower sensitivity of \emph{VISIR} in the SiC filter, we could not recenter the individual \emph{VISIR} images for the observations \#40 and \#44. The \#4 sequence was not included because of very low signal, maybe caused by some clouds, leading to a bad recentering of the individual images.

\addtocounter{table}{1}


\section{Data analysis}
\label{data_analysis}
In this section we first study the evolution of the atmospheric properties (seeing, transparency) during the observations for each night. We then present the aperture photometry we applied to our average images. We then combine these measurements with other photometric data we retrieve from the literature to study the spectral energy distribution of each star. Using a Fourier technique analysis on our average \emph{VISIR} images we also search for the presence of extended emission.

\subsection{Atmospheric evolution}

\begin{figure}[]
\resizebox{\hsize}{!}{\includegraphics[width=8.5cm]{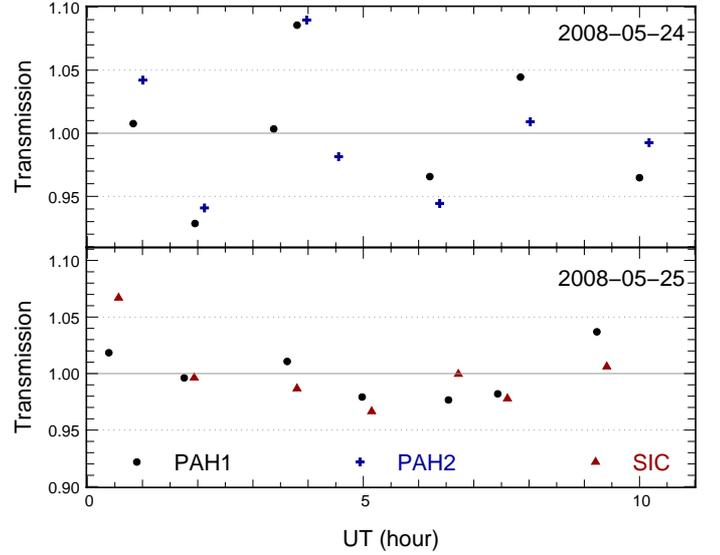}}
\caption{Relative atmospheric transmission for our observing nights. Both nights are normalised to the mean value over all calibrators observed. The dotted lines denote a 5\,\% variation.}
\label{transmission}
\end{figure}

We studied the sky transparency evolution using our photometric reference stars. The photometric templates from \citet{Cohen-1999-04} were used to plot the temporal relative atmospheric transmission (relative to the average value of all calibrators observed for each night). We see in Fig.~\ref{transmission} that the atmosphere was pretty stable during the observing nights. In the first night, the relative standard deviation is below 5\,\% while it is only 3\,\% in the second night.

\subsection{Photometry}
\label{photometry}

We carried out a classical aperture photometry to assess the flux density for each star in each filter. Photometric templates from \citet{Cohen-1999-04} were used to have an absolute calibration of the flux density taking into account the filter transmission\footnote{Filter transmission profiles are available on the ESO website http://www.eso.org/sci/facilities/paranal/instruments/visir/inst/index.html.} using:
\begin{displaymath}
F(\mathrm{ref}) = \frac{ \int_{\Delta\lambda} F_\lambda T_\lambda d\lambda }{ \int_{\Delta\lambda} T_\lambda d\lambda }
\end{displaymath}
where $F_\lambda$ is the reference irradiance (from the Cohen et al. templates), $T_\lambda$ the filter transmission and $\Delta\lambda$ the filter bandpass.

We applied an airmass correction factor:
\begin{displaymath}
F_\mathrm{corr} = F_\mathrm{obs} \times C(\lambda,AM)
\end{displaymath}
with $C(\lambda,AM)$ taken from \citet{Schutz-2005-}:
\begin{displaymath}
C(\lambda,AM) = 1 + \left[ 0.220 - \frac{0.104}{3}(\lambda - 8.6\,\mu\mathrm{m}) \right] (AM - 1)
\end{displaymath}

No aperture correction is required since we kept the same, relatively large, aperture radius (1.3\arcsec) for both the Cepheids and their respective reference stars. We then estimated the flux density of our Cepheid samples:
\begin{displaymath}
F(\mathrm{cep}) = \frac{F_\mathrm{corr}(\mathrm{cep})}{F_\mathrm{corr}(\mathrm{ref})}\,F(\mathrm{ref})
\end{displaymath}
where cep and ref stand for the Cepheid and reference stars.

The measured flux densities are summarized in Table~\ref{measured_flux_densities}. The uncertainties include the statistical dispersion of the aperture photometry, the dispersion of the calibrator flux densities over the night, and the absolute calibration uncertainty. Some final averaged images of our sample are presented in Fig.~\ref{VISIR_images}. The visibility of several diffraction rings reflects the quality of the data.

\begin{table*}[ht]
\centering
\caption{Measured flux densities of our Cepheid sample.}
\begin{tabular}{cccccc} 
\hline
\hline
Stars	  					&	    MJD			&	Filter			& 	Flux density 											& Flux density				& Excess 		\\
		  					&						&					&	($\mathrm{W/m^2/\mu m}$)				& 	(Jy)							& 	(\%)			\\
\hline
FF~Aql					& 	54~611.403	&  PAH1		&	$9.23\,\pm\, 0.25\,\times\,10^{-14}$	&	$2.27\,\pm\,0.06$	&	$1.6\,\pm\,2.8$	\\
							& 	54~611.411	&	SiC			&	$2.79\,\pm\, 0.08\,\times\,10^{-14}$	&	$1.28\,\pm\,0.06$	&	$3.7\,\pm\,3.0$	\\
\hline
AX~Cir					& 	54~611.076	&  PAH1		&	$6.97\,\pm\, 0.19\,\times\,10^{-14}$	&	$1.72\,\pm\,0.05$	&	$-0.8\,\pm\,2.9$	\\
\hline
X~Sgr					& 	54~610.104	&  PAH1		&	$2.31\,\pm\, 0.11\,\times\,10^{-13}$	&	$5.69\,\pm\,0.27$	&	$14.1\,\pm\,5.4$	\\
							&	54~611.112	&	PAH1		&	$2.11\,\pm\, 0.10\,\times\,10^{-13}$	&	$5.20\,\pm\,0.25$	&	$3.8\,\pm\,9.8$	\\
							&	54~610.111	&	PAH2		&	$7.82\,\pm\, 0.20\,\times\,10^{-14}$	&	$3.31\,\pm\,0.08$	&	$11.9\,\pm\,2.9$	\\
							&	54~611.119	&	SiC			&	$6.27\,\pm\, 0.15\,\times\,10^{-14}$	&	$2.89\,\pm\,0.07$	&	$5.6\,\pm\,8.1$	\\
\hline
$\eta$~Aql			&	54~610.282	&  PAH1		&	$3.96\,\pm\, 0.14\,\times\,10^{-13}$	&	$9.73\,\pm\,0.35$	&	$0.38\,\pm\,3.6$	\\
							&	54~611.248	&  PAH1		&	$4.30\,\pm\, 0.16\,\times\,10^{-13}$	&	$10.6\,\pm\,0.39$	&	$9.0\,\pm\,4.1$		\\
							&	54~610.289	&	PAH2		&	$1.39\,\pm\, 0.05\,\times\,10^{-13}$	&	$5.89\,\pm\,0.21$	&	$2.6\,\pm\,3.7$		\\
							&	54~611.255	&	SiC			&	$1.28\,\pm\, 0.05\,\times\,10^{-13}$	&	$5.90\,\pm\,0.23$	&	$9.6\,\pm\,4.3$		\\
\hline
W~Sgr					&	54~610.213	&  PAH1		&	$1.70\,\pm\, 0.07\,\times\,10^{-13}$	&	$4.19\,\pm\,0.17$	&	$10.4\,\pm\,4.6$ \\
							&	54~611.131	&  PAH1		&	$1.82\,\pm\, 0.05\,\times\,10^{-13}$	&	$4.48\,\pm\,0.12$	&	$18.2\,\pm\,3.3$	\\
							&	54~610.220	&	PAH2		&	$6.16\,\pm\, 0.36\,\times\,10^{-14}$	&	$2.61\,\pm\,0.15$	&	$16.5\,\pm\,6.8$	\\
							&	54~611.139	&	SiC			&	$5.30\,\pm\, 0.13\,\times\,10^{-13}$	&	$24.4\,\pm\,0.6$		&	$16.2\,\pm\,2.9$	\\
\hline
Y~Oph					&	54~610.126	&  PAH1		&	$2.01\,\pm\, 0.05\,\times\,10^{-13}$	&	$4.95\,\pm\,0.12$	&	$6.8\,\pm\,2.7$	\\
							&	54~611.170	&  PAH1		&	$2.00\,\pm\, 0.05\,\times\,10^{-13}$	&	$4.93\,\pm\,0.12$	&	$6.3\,\pm\,2.7$	\\
							&	54~610.134	&	PAH2		&	$6.62\,\pm\, 0.25\,\times\,10^{-14}$	&	$2.80\,\pm\,0.11$	&	$2.4\,\pm\,3.9$	\\
							&	54~611.177	&	SiC			&	$5.83\,\pm\, 0.15\,\times\,10^{-14}$	&	$2.69\,\pm\,0.07$	&	$4.6\,\pm\,2.7$	\\
\hline
U~Car					&	54~610.546	&  PAH1		&	$1.07\,\pm\, 0.02\,\times\,10^{-13}$	&	$2.64\,\pm\,0.05$	&	$32.1\,\pm\,2.5$	\\
							&	54~611.035	&	PAH1		&	$1.06\,\pm\, 0.05\,\times\,10^{-14}$	&	$2.61\,\pm\,0.12$	&	$30.9\,\pm\,6.2$	\\
							&	54~611.042	&	SiC			&	$2.91\,\pm\, 0.2\,\times\,10^{-14}$	&	$1.34\,\pm\,0.09$	&	$20.9\,\pm\,8.3$	\\
\hline
SV~Vul					&	54~611.365	&  PAH1		&	$8.62\,\pm\, 0.17\,\times\,10^{-14}$	&	$2.12\,\pm\,0.04$	&	$25.0\,\pm\,2.5$	\\
							&	54~611.372	&	SiC			&	$2.35\,\pm\, 0.07\,\times\,10^{-14}$	&	$1.08\,\pm\,0.03$	&	$15.1\,\pm\,3.4$	\\
\hline
R~Sct					&	54~610.236	&  PAH1		&	$6.74\,\pm\, 0.16\,\times\,10^{-13}$	&	$16.6\,\pm\,0.4$		&	$134\,\pm\,6$	\\
							&	54~611.189	&  PAH1		&	$7.05\,\pm\, 0.17\,\times\,10^{-13}$	&	$17.4\,\pm\,0.4$		&	$145\,\pm\,6$	\\
							&	54~610.243	&	PAH2		&	$2.56\,\pm\, 0.06\,\times\,10^{-13}$	&	$10.8\,\pm\,0.3$		&	$153\,\pm\,7$	\\
							&	54~611.196	&	SiC			&	$2.24\,\pm\, 0.06\,\times\,10^{-13}$	&	$10.3\,\pm\,0.3$		&	$158\,\pm\,7$	\\
\hline
AC~Her					&	54~610.349	&  PAH1		&	$9.18\,\pm\, 0.25\,\times\,10^{-13}$	&	$22.6\,\pm\,0.6$		&	$8273\,\pm\,228$	\\
							&	54~611.340	&  PAH1		&	$9.40\,\pm\, 0.23\,\times\,10^{-13}$	&	$23.2\,\pm\,0.6$		&	$8473\,\pm\,210$	\\
							&	54~610.357	&	PAH2		&	$10.1\,\pm\, 0.28\,\times\,10^{-13}$	&	$42.8\,\pm\,1.2$		&	$26705\,\pm\,743$	\\
							&	54~611.347	&	SiC			&	$8.33\,\pm\, 0.21\,\times\,10^{-13}$	&	$38.4\,\pm\,1.0$		&	$25586\,\pm\,647$	\\
\hline
$\kappa$~Pav		&	54~611.227	&  PAH1		&	$1.93\,\pm\, 0.05\,\times\,10^{-13}$	&	$4.75\,\pm\,0.12$	&	$22.2\,\pm\,3.2$	\\
							&	54~611.235	&	SiC			&	$5.39\,\pm\, 0.14\,\times\,10^{-14}$	&	$2.48\,\pm\,0.06$	&	$15.3\,\pm\,3.0$	\\
\hline
\end{tabular}
\tablefoot{The flux density was measured over an aperture radius of 1.3\arcsec. The parameter $\alpha$ is the measured flux density excess.}
\label{measured_flux_densities}
\end{table*}

\subsection{Spectral energy distribution of classical Cepheids}
\label{spectral_energy_distribution}

We have collected additional photometric measurements in the literature from $0.4\,\mu\mathrm{m}$ to $100\,\mu\mathrm{m}$ to analyse the spectral energy distribution of our Cepheids. As they are pulsating stars, the SEDs vary during the pulsation cycle and we have to take into account the phase of pulsation when retrieving the data.

\begin{figure}[!h]
\centering
\resizebox{\hsize}{!}{\includegraphics[width=4.45cm]{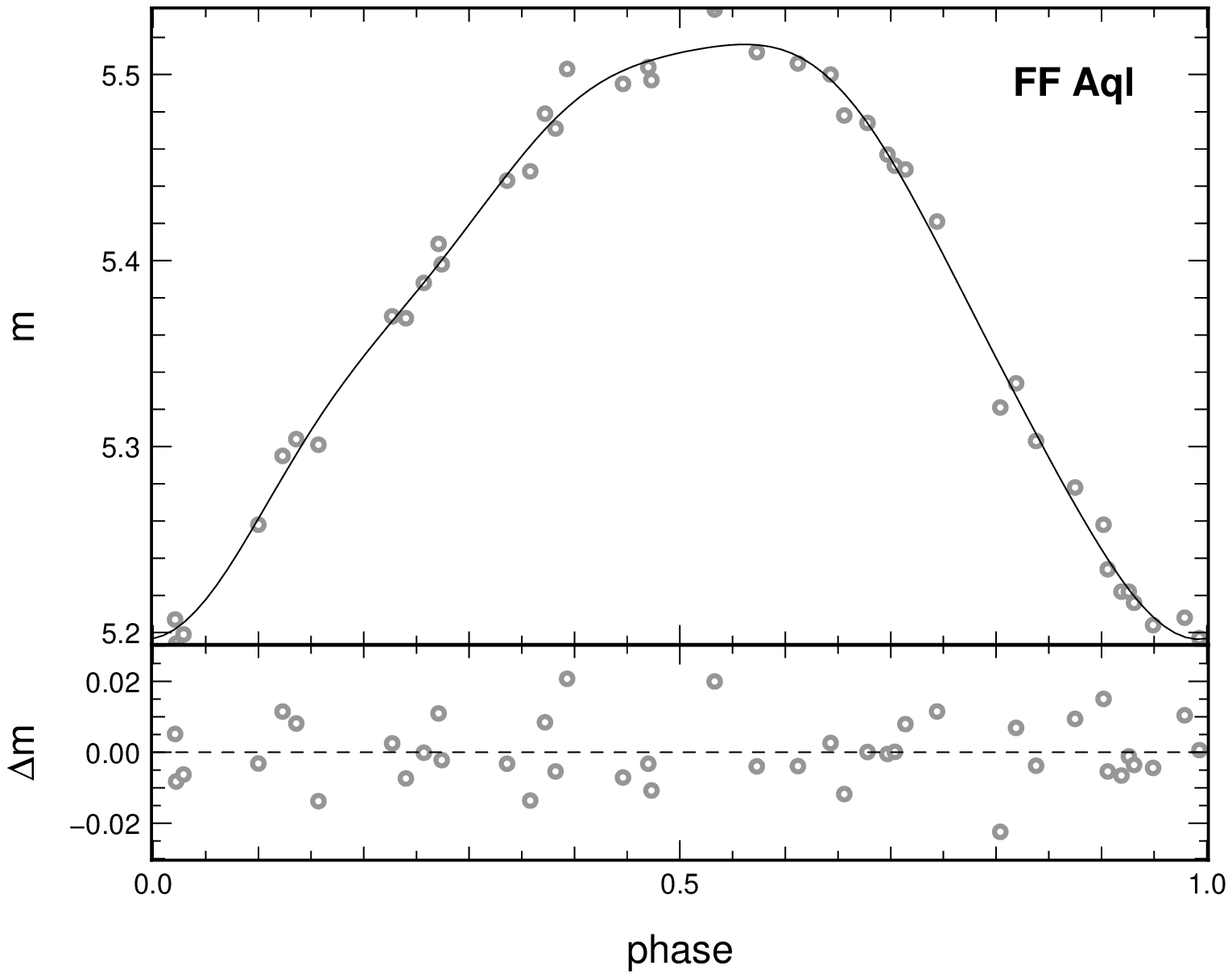}}
\resizebox{\hsize}{!}{\includegraphics[width=4.45cm]{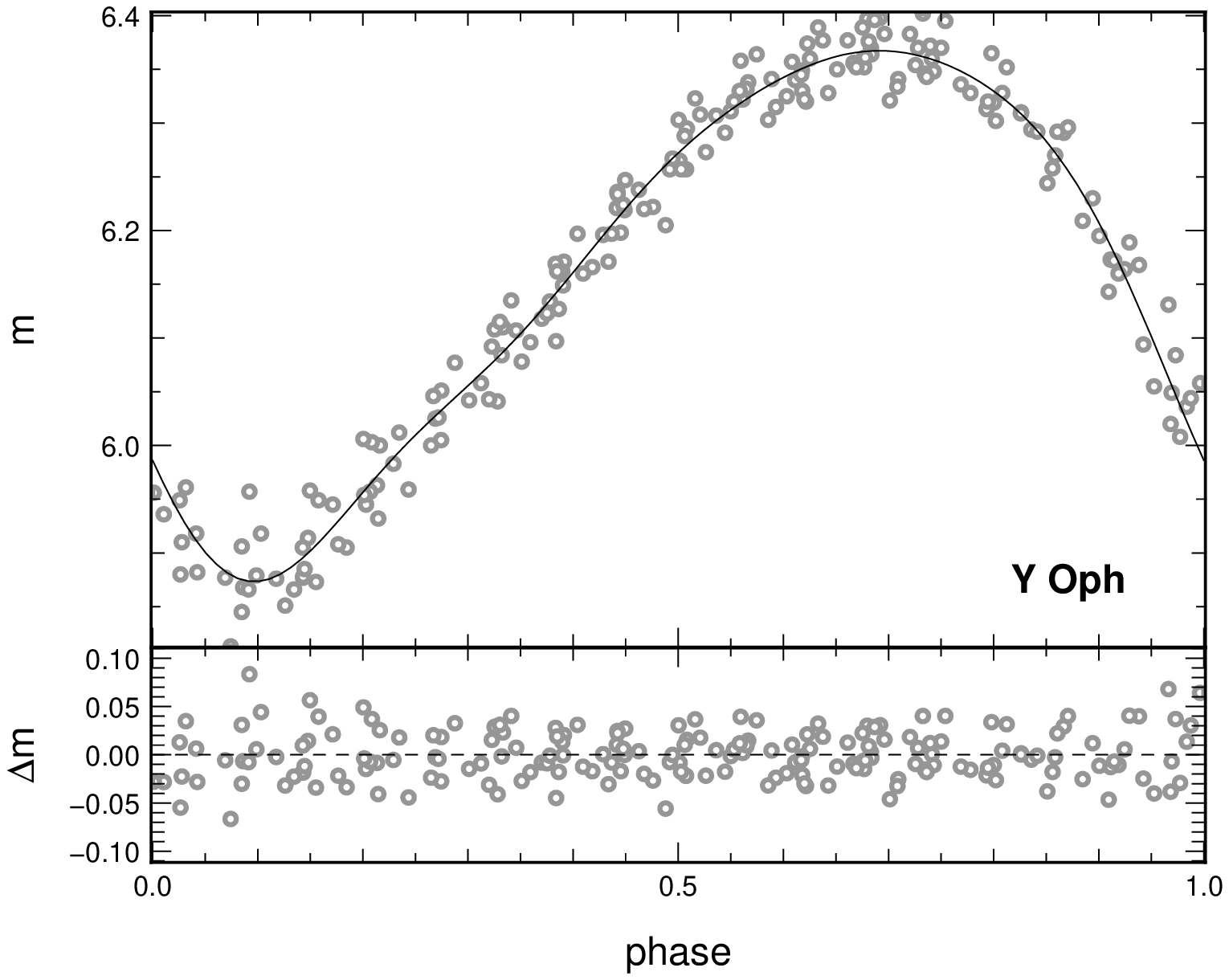}}
\caption{Examples for the constructed light curves in the $V$ band. Data for FF~Aql are from \citet{Moffett-1984-07} and from \citet{Berdnikov-2008-04} for Y~Oph. The resulting light curves are drawn with solid lines.}
\label{light_curve}
\end{figure}

To estimate the magnitudes at our phase of pulsation we retrieved light curves from the literature when available, that we plotted as a function of phase (all phases were computed with the ephemerides from Table~\ref{cepheid_parameter}). We then applied the Fourier decomposition technique \citep[see e.g.][]{Ngeow-2003-04}:

\begin{displaymath}
m = a_0 + \sum_{i=1}^na_i\,\cos(2\pi i\phi + b_i) 
\end{displaymath}
where $a_i$ and $b_i$ are the Fourier amplitudes and phases for the order $i$ to fit and $\phi$ is the pulsation phase. The parameter $n$ depends on the number of data points and on the light curve amplitude (higher amplitude require higher order fit). We used $n = 4$ for all stars. Examples are shown in Fig.~\ref{light_curve} for two stars. The coefficients of the fit were then used to compute values at our phases. We used this method for the Cepheids FF~Aql (in $B, V$ and $R$ bands), AX ~Cir ($B, V$), X Sgr ($B, V, J, H, K$), $\eta$~Aql ($B, V$), W~Sgr ($B, V$), Y~Oph ($B, V$), U~Car ($B, V$) and SV~Vul ($B, V$). When data points were not equally spaced in phase, the direct Fourier fit produced unrealistic oscillations, we so used a periodic cubic spline interpolation. To evaluate the uncertainty we used the total standard deviation of the residual values (bottom panels of Fig.~\ref{light_curve}). The uncertainty is considered the same for all phases. We used this kind of interpolation for the stars $\eta$~Aql (in $J, H, K$ bands), Y~Oph ($J, H, K$), U~Car ($J, H, K$), SV~Vul ($J, H, K$) and $\kappa$~Pav ($V, J, H, K$).

However there is a lack of light curves in the literature for some wavelengths and it is not possible to correct all the data for the phase mismatch. We then chose to use the amplitude of the light curves $A_\lambda$ as additional uncertainties on the magnitude due to the phase mismatch. In addition we know that $A_\lambda$ is decreasing with wavelength. Using 51 Galactic Cepheids, \citet{Laney-1993-01} plotted the amplitude of the $J, H$ and $K$ light curves as a function of the pulsation period (their Fig.~8). There are mainly two regions: $0.5<\log{P}\leqslant1.0$ with $A_J\,\sim\,0.1$\,mag and $\log{P}>1.0$ with $A_J\,\sim\,0.2$\,mag and these values decrease with the wavelength ($A_K\,\sim\,0.08$\,mag for $\log{P}\leqslant1.0$ and $A_K\,\sim\,0.15$\,mag for $\log{P}>1.0$). Therefore when the collected additional data do not correspond to same phases, we chose as additional uncertainties on the magnitude due to the phase mismatch the values 0.1\,mag if $0.5<\log{P}\leqslant1.0$ or  0.2\,mag if $\log{P}>1.0$ for $1<\lambda<3.5\,\mu$m and for $\lambda>3.5\,\mu$m we chose $\,0.05$\,mag for all periods. 
We are conscious that this can overestimate the uncertainties but this will enable us to completely separate an IR excess detection from a magnitude error on the SED curves. On the other hand this can also prevent the detection of a small IR excess.

Some of our targets were observed on two different nights and they can have different phases (especially the short periods). We so considered the SED according to an intermediate phase between the observations.

The photospheric emission was modelled with tabulated stellar atmosphere models obtained with the ATLAS9 simulation code from \citet{Castelli-2003-}. We have chosen a grid which was computed for solar metallicity and a turbulence velocity of $2\,\mathrm{km\,s^{-1}}$. We then interpolated this grid in order to compute spectra for any effective temperature and any surface gravity. The spectrum was multiplied by the solid angle of the stellar photosphere, $	\pi \theta_\mathrm{LD}^2/4$, where $\theta_\mathrm{LD}$ is the limb-darkened angular diameter. We adjusted the photometric data to the model taking into account the spectral response of each instrument. We assume that there is no detectable excess (5\,\% or less) below $2.2\,\mu\mathrm{m}$ and all the photometric measurements bluer than the $K$ band are used to fit the angular diameter and the effective temperature. 
However we should note that $T_\mathrm{eff}$ and $\theta_\mathrm{LD}$ are correlated variables in this fit. We did not adjust the surface gravity, since the broadband photometry is mostly insensitive to this parameter and its value was also chosen in the literature. 

All flux densities $< 3\,\mu$m are corrected for interstellar extinction $A_\lambda\,=\,R_\lambda\,E(B - V)$ using the total-to-selective absorption ratios $R_\lambda$ from \citet{Fouque-2003-} and \citet{Hindsley-1989-06}, and the color excess $E(B - V)$ from \citet{Fouque-2007-12}. Fluxes in any other longer wavelengths are not corrected for the interstellar extinction, which we assume to be negligible.

\subsubsection{FF~Aql}
\defcitealias{Luck-2008-07}{L08}
\defcitealias{Berdnikov-2008-04}{B08}
\defcitealias{Moffett-1984-07}{M84}
\defcitealias{Marengo-2010-01}{M10}

\begin{figure}[!h]
\centering
\resizebox{\hsize}{!}{\includegraphics{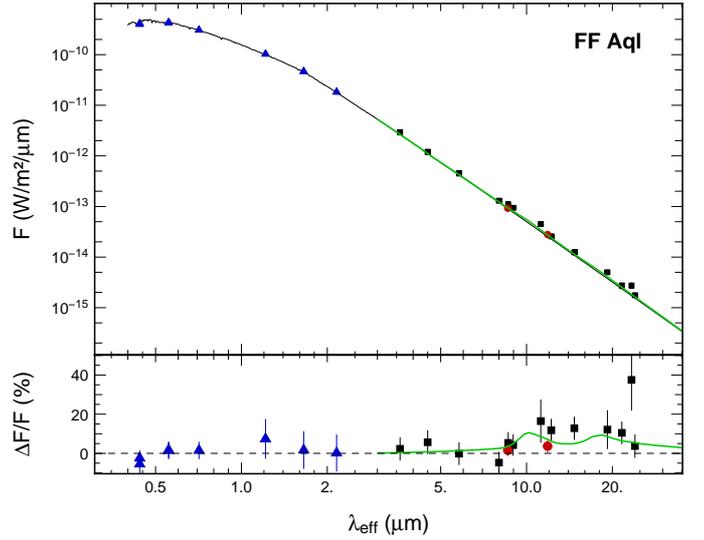}\hspace{.3cm}}
\caption{Synthetic spectra of the classical Cepheid FF~Aql (solid line) with the photometric measurements taken from the literature. The blue triangles are the points used to fit the SED. Our measurements are presented with the red circles while the black squares are the other photometric data. On the lower panel is plotted the excess flux density relatively to the photospheric emission. The green curve represents the flux density model.}
\label{SED_FF_AQL}
\end{figure}

We selected $\log{g} = 2.05$ from \citet[][hereafter \citetalias{Luck-2008-07}]{Luck-2008-07} as fixed parameter. The adjusted SED is presented in Fig.~\ref{SED_FF_AQL}. The $B$, $V$ and $R$ values were taken from light curves of \citet[][hereafter \citetalias{Berdnikov-2008-04}]{Berdnikov-2008-04} and \citet[][hereafter \citetalias{Moffett-1984-07}]{Moffett-1984-07}. The $J, H$ and $K_\mathrm{s}$ photometry are from \citet{Welch-1984-04} and correspond to the mean values (there was not a good coverage in phase to estimate accurate light curves, we so decided to take the mean values and their standard deviations). We also added photometric values from the Infrared Array Camera (\emph{IRAC}: 3.6, 4.5, 5.8 and 8\,$\mu$m) and the Multiband Imaging Photometer (\emph{MIPS}: 24\,$\mu$m) installed in the \emph{Spitzer} space telescope \citep[][ hereafter \citetalias{Marengo-2010-01}]{Marengo-2010-01}. We also use broadband photometry from the Infrared Astronomical Satellite \citep[\emph{IRAS}: 12 and 25\,$\mu$m,][]{Helou-1988-}, from the \emph{AKARI} satellite \emph{IRC} point source catalogue \citep[9 and 18\,$\mu$m,][]{Ishihara-2010-05} and from the Midcourse Space Experiment \citep[\emph{MSX}: 8.28, 12.13, 14.65, 21.34\,$\mu$m,][]{Egan-1996-12,Egan-2003-}.

0ur best-fit values are presented in Table~\ref{fitted_parameters} and plotted in Fig.~\ref{SED_FF_AQL} (black solid curve). This fitted effective temperature is only 3\,\% smaller than the value from \citetalias{Luck-2008-07} ($6062 \pm 43$\,K) at this phase of pulsation. The angular diameter is in excellent agreement with the $0.86 \pm 0.17$\,mas from \citet{Groenewegen-2007-11}. 

We detected a likely IR emission of the order of 2\,\% (with respect to the stellar flux) in PAH1 and SiC (see Table~\ref{measured_flux_densities}). FF~Aql presents a more important excess at other wavelengths (8 to 30\,$\mu$m). The lower panel in Fig.~\ref{SED_FF_AQL} represents the relative flux difference (relative to the model), integrated over the filter bandpass.

We fitted to this star a second component assuming an optically thin envelope that follows an intensity distribution of the form:

\defcitealias{Ossenkopf-1994-11}{O94}

\begin{equation}
F_\lambda = \kappa_\lambda\,\beta\,B_\lambda(T_\mathrm{d})
\label{equation__greybody}
\end{equation}
where $B_\lambda(T_\mathrm{d})$ is the Planck function at dust temperature $T_\mathrm{d}$, $\kappa_\lambda$ is the dust opacity and $\beta$ is a parameter directly proportional to the dust mass \citep{Li-2005-04}:
\begin{displaymath}
\beta = 2.1\times10^{-3}\,\frac{M_\mathrm{d}}{D^2}
\end{displaymath}
with $D$ the distance of the star in pc and $M_\mathrm{d}$ in $M_\odot$.

The IR excess seems to rise quite sharply around $10\,\mu\mathrm{m}$ and $20\,\mu\mathrm{m}$ that could be linked to silicate/carbon features.  We so chose the dust opacity from \citet[hereafter \citetalias{Ossenkopf-1994-11}]{Ossenkopf-1994-11} for carbon and silicate grains for a MRN size distribution \citep{Mathis-1977-10}. 


We plotted in Fig.~\ref{SED_FF_AQL} (the green solid curve) our best flux density model with a temperature $T_\mathrm{d} = 539 \pm 52$\,K and $\beta = 3.0 \pm 0.7 \times 10^{-19}\,\mathrm{kg\,m^{-2}}$. Using the distance from \citet[][Table~\ref{cepheid_parameter}]{Benedict-2007-04}, this leads to a dust mass of $M_\mathrm{d} = 1.8 \pm 0.4 \times 10^{-11}\,M_\odot$ and to a total mass (gas + dust) of $1.8 \pm 0.4 \times 10^{-9}\,M_\odot$ (using a gas to dust ratio of $\sim 100$ typical of circumstellar dust). We have to mention that some of the additional broadband photometric values could be overestimated due to the surrounding interstellar cirrus emission \citep{Barmby-2010-11}. 

\begin{table*}[ht]
\centering
\caption{Best-fit values for our Cepheid sample.}
\begin{tabular}{cccccccc} 
\hline
\hline
Stars	  				&	$\phi$	&	$\log g$	&	$\theta_\mathrm{LD}$	& 	$T_\mathrm{eff}$	& $M_\mathrm{cse}$	& $T_\mathrm{cse}$	&	$\chi^2$	\\
		  				&				&					&				(mas)				&	(K)						& 	($M_\odot$)			& 	(K)						&					\\
\hline
FF~Aql				&	0.62		&	2.05	&	$0.86 \pm 0.03$	&	$5890 \pm 235$	&	$ 1.8 \pm 0.5 \times 10^{-9}$	&	$539 \pm 53$		&	0.41	\\
AX~Cir				& 	0.27		&	2.00	&	$0.76 \pm 0.03$	&	$5911 \pm 184$	&	$ 7.4 \pm 5.9 \times 10^{-10}$&	$712 \pm 61$		&	0.42	\\
X~Sgr				& 	0.72		&	2.00	&	$1.30 \pm 0.04$	&	$5738 \pm 314$	&	$ 3.0 \pm 0.6 \times 10^{-9}$	&	$703 \pm 52$		&	0.74	\\
$\eta$~Aql		&	0.47		&	1.80	&	$1.86 \pm 0.12$	&	$5431 \pm 498$	&	$ 1.0 \pm 0.7 \times 10^{-8}$	&	$545 \pm 60$		&	1.01	\\
W~Sgr				&	0.48		&	1.70	&	$1.14 \pm 0.06$	&	$5632 \pm 162$	&	$ 1.7 \pm 0.4 \times 10^{-9}$	&	$853 \pm 55$		&	1.45	\\
Y~Oph				&	0.73		&	1.80	&	$1.24 \pm 0.05$	&	$5870 \pm 387$	&	$ 3.9 \pm 0.8 \times 10^{-9}$	&	$1419 \pm 148$	&	1.49	\\
U~Car				&	0.49		&	1.20	&	$0.90 \pm 0.02$	&	$4823 \pm 52$		&	$ 4.4 \pm 2.1 \times 10^{-9}$	&	$746 \pm 94$		&	0.70	\\
SV~Vul				&	0.04		&	1.40	&	$0.76 \pm 0.01$	&	$5744 \pm 144$	&	$ 3.9 \pm 7.4 \times 10^{-8}$	&	$620 \pm 50$		&	1.13 \\
\hline
R~Sct				&	0.48		&	0.00	&	$1.74 \pm 0.06$	&	$4605 \pm 119$	&	--												&	$1486 \pm 335$	&	1.79	\\
						&				&			&								&								&	--												&	$772 \pm 82$		&			\\
AC~Her				&	0.14		&	0.50	&	$0.30 \pm 0.02$ 	&	$5711 \pm 293$	&	--												&	$286 \pm 32$		&	0.91	\\
$\kappa$~Pav	&	0.90		&	1.20	&	$1.09 \pm 0.05$	&	$6237 \pm 119$	&	$9.9 \pm 2.5 \times 10^{-10}$	&	$695 \pm 36$		&	0.30	\\
\hline
\end{tabular}
\tablefoot{$\phi$ is the pulsation phase. $\theta_\mathrm{LD}$ and $T_\mathrm{eff}$ are the photospheric fitted parameters while $T_\mathrm{cse}$ is the CSE fitted parameter. $M_\mathrm{cse}$ is the total mass (dust + gas) of the CSE estimated using a gas to dust ratio of $\sim 100$. For R~Sct, the first line denotes the compact component. The $\chi^2$ has been estimated from the fit of the photospheric flux. See the text for the references of the effective gravity.}
\label{fitted_parameters}
\end{table*}

\subsubsection{AX~Cir}
\defcitealias{Moskalik-2005-06}{M05}

\begin{figure}[!h]
\centering
\resizebox{\hsize}{!}{\includegraphics{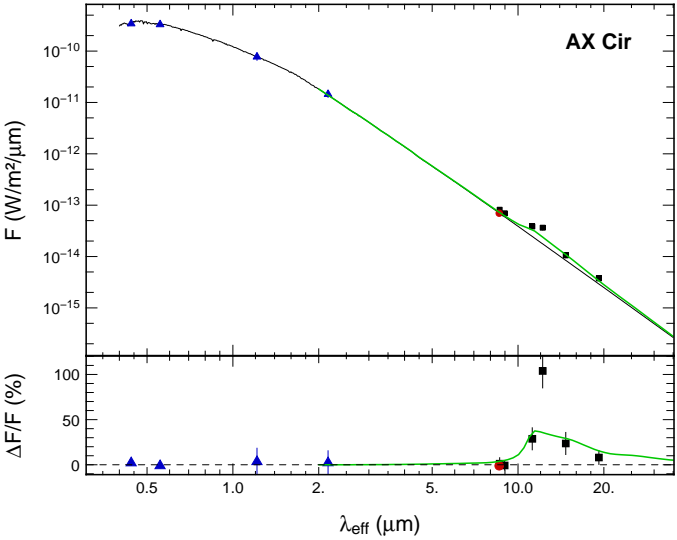}\hspace{.3cm}}
\caption{The same as Fig.~\ref{SED_FF_AQL}, but for AX~Cir.}
\label{SED_AX_CIR}
\end{figure}

As AX~Cir has almost the same spectral type than FF~Aql we considered a surface gravity of $\log{g} = 2.0$ (note that a change of $\pm 0.5$ only change our following estimated values by 0.7\,\%). Fig.~\ref{SED_AX_CIR} shows the SED fit for this Cepheid. $B$ and $V$ photometry are from \citetalias{Berdnikov-2008-04} (from light curves) with a color excess index from \citet{Tammann-2003-06}. Data were also retrieved from the Deep Near Infrared Survey of the Southern Sky (\emph{DENIS}) for the $J$ and $K_\mathrm{s}$ bands. Beyond 9\,$\mu$m data come from \emph{MSX}, \emph{IRC} and \emph{IRAS}.

Our fitted parameters are presented in Table~\ref{fitted_parameters}. The diameter is in agreement with the mean value predicted by \citet[][hereafter \citetalias{Moskalik-2005-06}, at a 7\,\% level, consistent with the angular diameter variation]{Moskalik-2005-06}.

We did not detect IR emission at 8.6\,$\mu$m and the value is consistent with the 9\,$\mu$m measurement from \emph{IRC} (see Table~\ref{measured_flux_densities}). However, it seems to have an infrared excess at longer wavelengths. We notice that the IR excess seems to rise around $13\,\mu\mathrm{m}$ and could be a feature of alumina oxide emission. Using optical constant for amorphous compact alumina \citep{Begemann-1997-02}, we computed $\kappa_\lambda$ from a simple model of homogeneous spheres in the Rayleigh limit of small particles \citep{Bohren-1983-}, with a size distribution from \citet[][with particle sizes from 5\,nm to 250\,nm as \citetalias{Ossenkopf-1994-11}]{Mathis-1977-10} and a density $\rho = 2.5\,\mathrm{g\,cm^{-2}}$. We then used this dust opacity in the model of Eq.~\ref{equation__greybody} to fit a second component (green curve in Fig.~\ref{SED_AX_CIR}), assuming that the envelope is mainly composed of amorphous alumina. We obtained $\beta = 1.6 \pm 0.4 \times 10^{-19}\,\mathrm{kg\,m^{-2}}$. Table~\ref{fitted_parameters} shows the fitted temperature and the total mass of the CSE estimated using a gas to dust ratio of $\sim 100$. However, a larger photometric dataset is necessary to better constrain the CSE parameters and its dust composition.

\subsubsection{X~Sgr}
\defcitealias{Kervella-2004-03}{K04}
\defcitealias{Feast-2008-06-2}{F08}

\begin{figure}[!h]
\centering
\resizebox{\hsize}{!}{\includegraphics{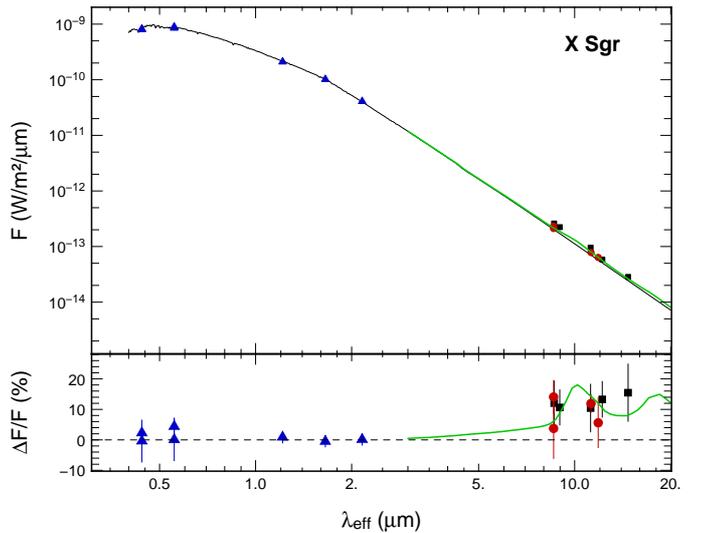}\hspace{.3cm}}
\caption{The same as Fig.~\ref{SED_FF_AQL}, but for X~Sgr.}
\label{SED_X_SGR}
\end{figure}

We retrieve photometric data from the $B, V$ dereddened magnitudes of \citet[][hereafter \citetalias{Kervella-2004-03}]{Kervella-2004-03} and from \citetalias{Moffett-1984-07} (from light curves at an intermediate phase $\phi = 0.72$). The other photometric data come from \citet[][from $J, H, K$ light curves, hereafter \citetalias{Feast-2008-06-2}]{Feast-2008-06-2}, \emph{IRC} (9\,$\mu$m), \emph{IRAS} (12\,$\mu$m) and \emph{MSX} (8.28, 12.13 and 14.65\,$\mu$m). These data are plotted in Fig.~\ref{SED_X_SGR}.
 
For the same surface gravity previously used ($\log{g} = 2.0$) we found an angular diameter (see Table~\ref{fitted_parameters}) that is 13\% and $3\sigma$ smaller than the mean diameter measured by \citet[][$1.47 \pm 0.04$\,mas, see Table~\ref{cepheid_parameter}]{Kervella-2004-03a}. However, the amplitude of the pulsation is $\sim 9$\,\% in diameter \citepalias{Moskalik-2005-06}. Conversely our estimate is in agreement with $\theta_\mathrm{LD} = 1.31 \pm 0.12$\,mas assessed from the parallax \citep{Benedict-2007-04} and the linear diameter at this phase of pulsation \citepalias{Feast-2008-06-2}. However \citetalias{Kervella-2004-03} used a limb darkened diameter to model their data and the presence of an extended emission could overestimate the angular diameter.

We detected a likely IR emission of the order of 5--15\,\% (with respect to the stellar flux, see Table~\ref{measured_flux_densities}). We fitted the same flux density model previously used, with the dust opacity from \citetalias{Ossenkopf-1994-11} as it seems to have some silicate/carbon features. We obtained $\beta = 5.6 \pm 0.9 \times 10^{-19}\,\mathrm{kg\,m^{-2}}$. The other parameters are presented in Table~\ref{fitted_parameters}. 

\subsubsection{$\eta$~Aql}

\begin{figure}[!h]
\centering
\resizebox{\hsize}{!}{\includegraphics{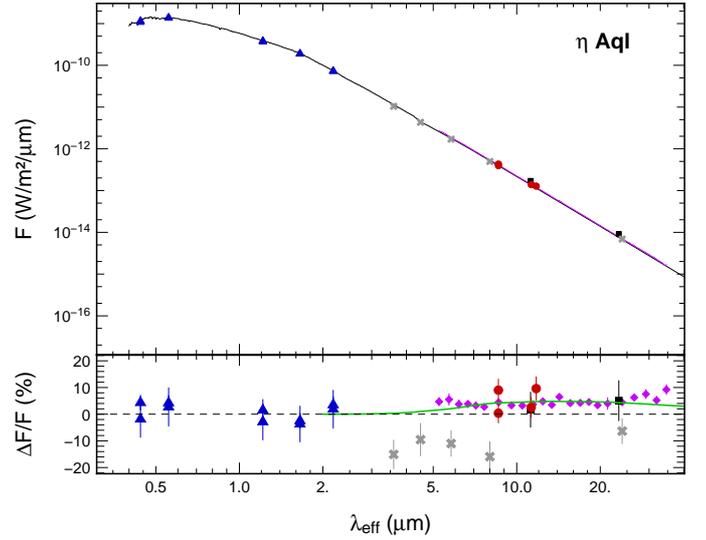}\hspace{.3cm}}
\caption{The same as Fig.~\ref{SED_FF_AQL}, but for $\eta$~Aql. The measured spectra were superimposed in purple.}
\label{SED_ETA_AQL}
\end{figure}

Optical and near-infrared photometry from \citetalias{Kervella-2004-03} ($B, V, J, H, K$), \citetalias{Moffett-1984-07} ($B, V$) and \citet[$J, H, K$,][]{Barnes-1997-06} were used for the SED (from light curves at an intermediate phase $\phi = 0.47$). For the longer wavelengths we retrieve the \emph{IRAS} fluxes and the \emph{Spitzer/IRS} spectra \citep[from 5 to 35\,$\mu$m,][]{Ardila-2010-12}. The result is plotted on Fig.~\ref{SED_ETA_AQL}. The \emph{IRS} spectra is plotted in purple. This spectra was acquired at a phase $\phi = 0.45$ very close to our intermediate phase of our observations and therefore we did not add uncertainty due to the phase mismatch. In the lower panel of Fig.~\ref{SED_ETA_AQL} we just plotted the spectra for some value (bin of 10\,$\mu$m) for clarity. The \emph{Spitzer} flux measurements from \citetalias{Marengo-2010-01} were also added, but are significantly lower than the stellar SED derived from visible and near-IR. A possible explanation of this inconsistency could be due to the different observing mode used for $\eta$~Aql with respect to the other Cepheids. $\eta$~Aql was observed in full-frame mode while the other stars used in this paper were observed in subarray mode. In the full frame mode the images of this bright star are heavily saturated. The PSF-fitting method used in \citetalias{Marengo-2010-01} could be incorrect, leading to an underestimation of the photometry. For this reason, we decided not to include them in the fit.

We found $\theta_\mathrm{LD}$ (see Table~\ref{fitted_parameters}) in good agreement with the measured diameter from \citet[][$\theta_\mathrm{LD} = 1.87 \pm 0.03$\,mas]{Kervella-2004-03a} at this phase and also with the mean angular diameter from \citet[][$\theta_\mathrm{LD} = 1.76 \pm 0.09$\,mas]{Groenewegen-2008-09}. At an intermediate phase ($\phi \sim 0.47$) \citet{Luck-2004-07} give a surface gravity of $\log{g} \sim 1.8$ and an effective temperature of about $T_\mathrm{eff} = 5508 \pm 40$\,K that is only 1\,\% larger than our estimated value.

We can see from the \emph{IRS} spectra that it is likely that this Cepheid has a small IR emission, starting with a relative excess (still with respect to the photosphere) of $4.7\,\pm\,1.5\,\%$ at 5.3\,$\mu$m to $9.2\,\pm\,1.8\,\%$ at 34.7\,$\mu$m. We also detect in our \emph{VISIR} filters an excess with the same order of magnitude (see Table~\ref{measured_flux_densities}).

No strong features of silicate seem to be present in the spectra. We so made the hypothesis that the envelope is mainly composed of amorphous carbon. Using optical constant for amorphous carbon \citep{Preibisch-1993-11}, we also computed $\kappa_\lambda$ from a simple model of homogeneous spheres in the Rayleigh limit of small particles, with a size distribution from \citet[][with particle sizes from 5\,nm to 250\,nm as \citetalias{Ossenkopf-1994-11}]{Mathis-1977-10} and a density $\rho = 2.0\,\mathrm{g\,cm^{-2}}$. We then used this dust opacity in Eq.~\ref{equation__greybody} to fit the infrared excess (green curve in Fig.~\ref{SED_ETA_AQL}). We obtained $\beta = 1.6 \pm 0.4\times 10^{-18}\,\mathrm{kg\,m^{-2}}$ that gives the total dust mass presented in Table~\ref{fitted_parameters} with the dust temperature.

\subsubsection{W~Sgr}

\begin{figure}[!h]
\centering
\resizebox{\hsize}{!}{\includegraphics{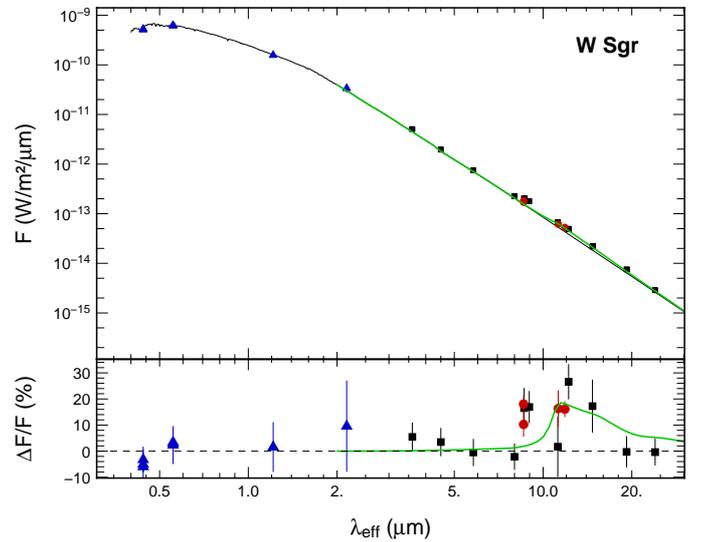}\hspace{.3cm}}
\caption{The same as Fig.~\ref{SED_FF_AQL}, but for W~Sgr.}
\label{SED_W_SGR}
\end{figure}

The $B$ and $V$ photometry are from \citetalias{Kervella-2004-03} and \citetalias{Berdnikov-2008-04} (at an intermediate phase $\phi = 0.48$). The other irradiances are from \emph{DENIS} ($J, Ks$), \emph{Spitzer} (3.6, 4.5, 5.8, 8 and 24\,$\mu$m from \citetalias{Marengo-2010-01}), \emph{IRC} (9 and 18\,$\mu$m), \emph{MSX} (8.28, 12.13, 14.65, 21.34\,$\mu$m) and the 12\,$\mu$m photometry from \emph{IRAS}. The SED is plotted in Fig.~\ref{SED_W_SGR}.

Our fitted effective temperature (presented in Table~\ref{fitted_parameters}) is in agreement at a 2\,\% level with $T_\mathrm{eff} = 5535 \pm 51$\,K from \citet{Luck-2004-07} around the same phase with an effective gravity from the same author of 1.7. We found an angular diameter that is about 15\,\% and $2\sigma$ smaller than the one measured by interferometry by \citetalias{Kervella-2004-03} ($1.31 \pm 0.04$\,mas) at this phase of pulsation. Conversely our estimate is in agreement with the mean diameter from \citet{Bersier-1997-04} who found $\overline{\theta_\mathrm{LD}} = 1.17 \pm 0.11$\,mas based on photometry. However \citetalias{Kervella-2004-03} used a limb darkened diameter to model their data and the presence of an extended emission could overestimate the angular diameter.

\emph{Spitzer}'s values are consistent with the blackbody radiation adjusted by \citetalias{Marengo-2010-01}. However our \emph{VISIR} photometry shows an excess of  $\pm\,15$\,\% (see Table~\ref{measured_flux_densities}). The same trend is observed from IRC (at 9\,$\mu$m) and \emph{MSX} (at 8.28, 12.13 and 14.65\,$\mu$m). This could be an evidence of a particular dust composition. As no IR emission is detected around $20\,\mu\mathrm{m}$, we rejected a silicate dust composition and we so assumed an envelope mainly composed of alumina oxide. We used the same dust opacity model as AX~Cir in Eq.~\ref{equation__greybody}. The resulting curve is plotted in Fig.~\ref{SED_W_SGR} and the fitted parameters are listed in Table~\ref{fitted_parameters}.


\subsubsection{Y~Oph}
\begin{figure}[!h]
\centering
\resizebox{\hsize}{!}{\includegraphics{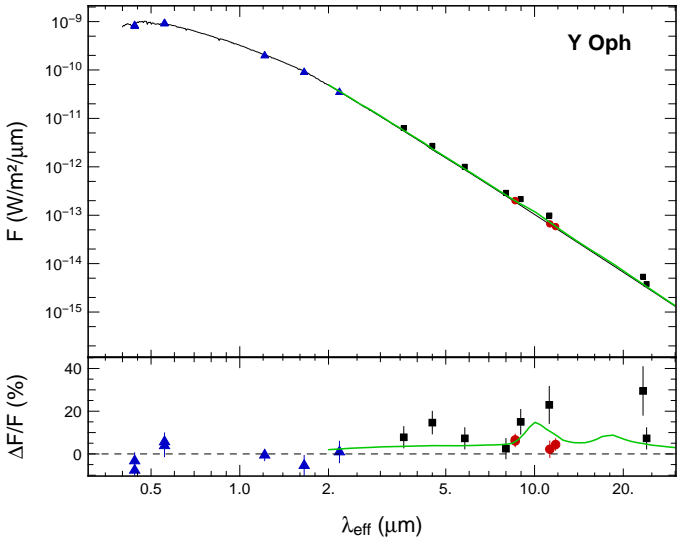}\hspace{.3cm}}
\caption{The same as Fig.~\ref{SED_FF_AQL}, but for Y~Oph.}
\label{SED_Y_OPH}
\end{figure}

We retrieved the $B, V$ data from \citetalias{Berdnikov-2008-04} and \citetalias{Moffett-1984-07}. The $J, H, K$ photometry is from \citet{Laney-1992-04} (also from light curves at an intermediate phase $\phi = 0.73$) and \citetalias{Kervella-2004-03}. The plot of the spectral energy distribution is shown in Fig.~\ref{SED_Y_OPH}. The other fluxes are from \emph{Spitzer} (3.6, 4.5, 5.8, 8 and 24\,$\mu$m from \citetalias{Marengo-2010-01}), \emph{IRC} (9\,$\mu$m) and the 12\,$\mu$m and 25\,$\mu$m photometry from \emph{IRAS}.

The best fitted parameters are presented in Table~\ref{fitted_parameters} for a found $\log{g} \sim 1.8$  \citepalias{Luck-2008-07}. \citetalias{Luck-2008-07} also give an effective temperature at this intermediate phase of $T_\mathrm{eff} = 5800 \pm 148$\,K that is in good agreement with our value. Our fitted angular diameter and the value $\theta_\mathrm{LD} = 1.24 \pm 0.01$\,mas at $\phi \sim 0.73$ measured by \citet{Merand-2007-08} is also good.

We detected a likely infrared emission from our photometric measurements (see Table~\ref{measured_flux_densities}). This result is consistent with \citet{Merand-2007-08} where a CSE has been found around this star in the $K$ band with a relative contribution of $5.0 \pm 2.0\,\%$. As some features of silicate/carbon seems to be present around $10\,\mu\mathrm{m}$ and $20\,\mu\mathrm{m}$, the model of Eq.~\ref{equation__greybody} was fitted with the dust opacity from \citetalias{Ossenkopf-1994-11}. The best-fitted parameters are presented in Table~\ref{fitted_parameters}. Y~Oph seems to have a hotter circumstellar envelope that could be located close to the star and heated by its radiations.


\subsubsection{U~Car}

\begin{figure}[!h]
\centering
\resizebox{\hsize}{!}{\includegraphics{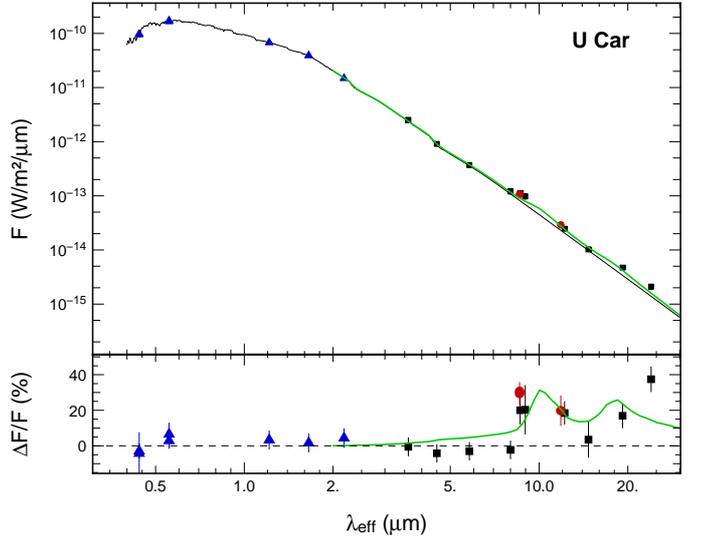}\hspace{.3cm}}
\caption{The same as Fig.~\ref{SED_FF_AQL}, but for U~Car.}
\label{SED_U_CAR}
\end{figure}

We selected the $B, V$ photometry from \citetalias{Berdnikov-2008-04} and \citet{Coulson-1985-} (from light curves at an intermediate phase $\phi = 0.63$). We also used the $J, H, K$ values from \citet[][hereafter L92, also from light curves]{Laney-1992-04} and mid-IR photometric measurements from \emph{IRC} (9 and 18\,$\mu$m), from \emph{MSX} (8.28, 12.13, 14.65\,$\mu$m) and \emph{Spitzer} (3.6, 4.5, 5.8, 8 and 24\,$\mu$m from \citetalias{Marengo-2010-01}). In Fig.~\ref{SED_U_CAR} is plotted the SED.

Our fitted angular diameter and an effective temperature with a given $\log{g} \sim 1.2$ from \citet{Romaniello-2008-09} (given at a phase of 0.49 but as said previously the broadband photometry is almost insensitive to the effective gravity changes) are listed in Table~\ref{fitted_parameters}. Our estimate of the diameter is in agreement at a 5\,\% level with the mean angular diameter from \citet[][$0.94 \pm 0.05$\,mas]{Groenewegen-2008-09}. However our estimated effective temperature is not consistent with the mean value $T_\mathrm{eff} = 5980$\,K estimated from the iron abundances by \citet{Romaniello-2008-09}. From a color-temperature relation \citep{Fry-1999-10}, the temperature of the star should be $T_\mathrm{eff} \sim 5000$\,K making suspicious the value from \citet{Romaniello-2008-09}. We did not find other measured temperatures in the literature to verify our estimate but we are confident in our value since all photometric data in the SED ($< 8\,\mu$m) look consistent with this effective temperature.

We have detected with \emph{VISIR} a significant IR emission (see Table~\ref{measured_flux_densities}). This is also visible at larger wavelength with an infrared excess of about 20--30\,\%. We also fitted the flux density model (Eq.~\ref{equation__greybody}) with the dust opacity from \citetalias{Ossenkopf-1994-11}. The results are listed in Table~\ref{fitted_parameters}.
We should notice that these values could be affected by the interstellar cirrus background emission in the region of this star \citep{Barmby-2010-11}.


\subsubsection{SV~Vul}

\begin{figure}[!h]
\centering
\resizebox{\hsize}{!}{\includegraphics{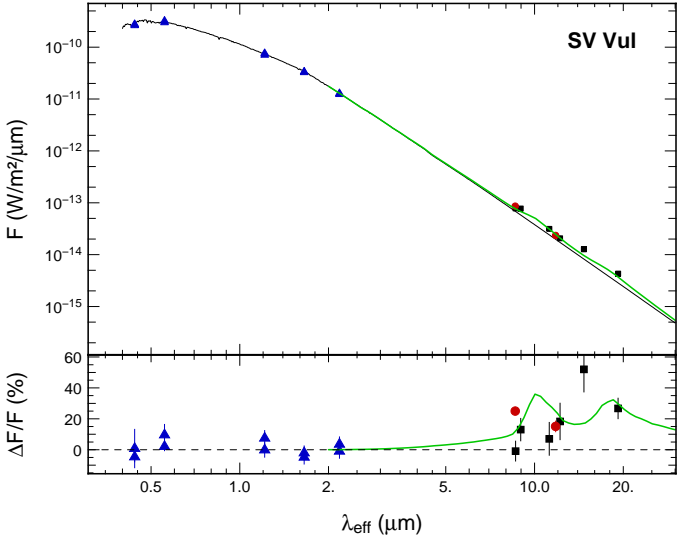}\hspace{.3cm}}
\caption{The same as Fig.~\ref{SED_FF_AQL}, but for SV~Vul.}
\label{SED_SV_VUL}
\end{figure}

The $B, V$ photometry were retrieved from \citetalias{Berdnikov-2008-04} and \citetalias{Moffett-1984-07} while the $J, H, K$ magnitudes from light curves of \citet{Barnes-1997-06} and \citet{Laney-1992-04}. Other infrared data are from \emph{IRC} (9 and 18\,$\mu$m), \emph{MSX} (8.28, 12.13, 14.65\,$\mu$m) and \emph{IRAS} (12\,$\mu$m). The spectral energy distribution is presented in Fig.~\ref{SED_SV_VUL}.

We took $\log{g} \sim 1.4$ from \citet{Kovtyukh-2005-01}. These authors give an effective temperature around the same phase of $T_\mathrm{eff} = 5977 \pm 32$\,K. We found a temperature that is only 4\,\% and $1.4\sigma$ smaller. The angular diameter we found is 5\,\% smaller than the mean value from \citet[][$0.80 \pm 0.05$\,mas]{Groenewegen-2008-09}, but within $1\sigma$.

We have detected an excess in PAH1 and in SiC filters (see Table~\ref{measured_flux_densities}). This excess is also visible at other wavelength up to $26.7\,\pm\,6.6\,\%$ at 18\,$\mu$m with \emph{IRC}. This indicates the presence of circumstellar material.

We fitted the same model previously used (Eq.~\ref{equation__greybody}) for the wavelengths larger than 3\,$\mu$m and still with the dust opacity from \citetalias{Ossenkopf-1994-11}. The fitted parameters are listed in Table~\ref{fitted_parameters}. The uncertainty in the total mass estimate of the envelope is due to the uncertainty on the parallax measurement (Table~\ref{cepheid_parameter}). More photometric data points are necessary to better constraint this SED, measurements at least between 3 and 8\,$\mu$m should be useful.


\subsection{Spectral energy distribution of type II Cepheids}

In this section we use the same methods and models for type II Cepheids than for classical Cepheids.

\subsubsection{R~Sct}

\begin{figure}[!h]
\centering
\resizebox{\hsize}{!}{\includegraphics{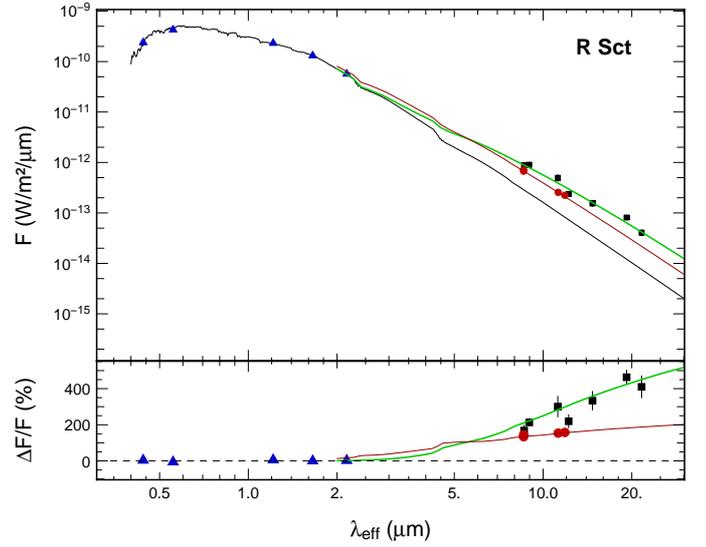}\hspace{.3cm}}
\caption{Synthetic spectra of the type II Cepheid R~Sct (solid line) with the photometric measurements taken from the literature. In blue are the points used to fit the SED. Our measurements are presented in red. On the lower panel is plotted the excess flux density relatively to the photospheric emission. The red curve represents the flux density model for a compact component while the green curve denotes larger one.}
\label{SED_R_SCT}
\end{figure}

This very irregular pulsator with a large amplitude has been extensively observed and is known to have an infrared excess \citep[see e.g.][]{Goldsmith-1987-07,Giridhar-2000-03,de-Ruyter-2005-05}. Its irregular pulsation and its large amplitude make it a special case as RV Tauri stars. The $B, V, J, H, K$ photometry were retrieved from \citet{Shenton-1994-07} (at phase 0.48). We also added values from \emph{IRC} (9 and 18\,$\mu$m), \emph{IRAS} (12\,$\mu$m) and from \emph{MSX} (8.28, 12.13, 14.65, 21.34\,$\mu$m). E($B$-$V$) and $B$-$V$ were obtained from \citet{Taranova-2010-02} and \citet{Myers-2001-09} respectively. The SED is plotted in Fig.~\ref{SED_R_SCT}.

We fitted the stellar component (i.e. $\lambda < 3\,\mu$m) with an effective gravity of $\log{g} = 0.0$ given by \citet{Giridhar-2000-03}. As expected we also detected a strong IR emission at our wavelengths (see Table~\ref{measured_flux_densities}). The fitted photospheric curve is shown in Fig.~\ref{SED_R_SCT} (solid black curve) and its parameters in Table~\ref{fitted_parameters}. Our temperature is consistent at a 2\,\% level with \citet{Giridhar-2000-03} who gave $T_\mathrm{eff} = 4500$\,K (at $\phi = 0.44$ using our ephemeris). \citet{Shenton-1994-07} estimated a diameter of about $1.79 \pm 0.59$\,mas (taking $d = 431\,$pc from \emph{Hipparcos}) and our value is only 3\,\% smaller. 

For the circumstellar component, we assumed that the envelope is optically thick and the observed flux density can be fitted by the following equation:
\begin{equation}
F_\lambda = A\,B_\lambda(T_\mathrm{d})
\label{equation__blackbody}
\end{equation}
where $A$ is related to the solid angle and the emissivity of the envelope.

By looking at the SED our values seem to be lower than the other measurements and it seems to have two different intensity distributions. As the effective resolution of \emph{VISIR} is larger than that of the \emph{AKARI} and \emph{IRAS} telescopes in the $N$ band, \emph{VISIR} only detected a compact component located close to the star. We so decided to first fit only our values as a compact component and then to fit a larger component using only the IR data from \emph{AKARI} and \emph{IRAS}. The result are presented in Fig.~\ref{SED_R_SCT}. The solid red curve is the possible warmer compact component for which our best fitted values were $T = 1486 \pm 335$\,K and $A = 4.65 \pm 0.83$\,mas$^2$. For the larger component (the green curve) we found $T = 772 \pm 83$\,K and $A = 11.2 \pm 1.25$\,mas$^2$ (Table~\ref{fitted_parameters}). This latter temperature is in the same range than the dust temperature $T = 800 \pm 50$\,K modelled by \citet{Taranova-2010-02}.

\subsubsection{AC~Her}
\begin{figure}[!h]
\centering
\resizebox{\hsize}{!}{\includegraphics{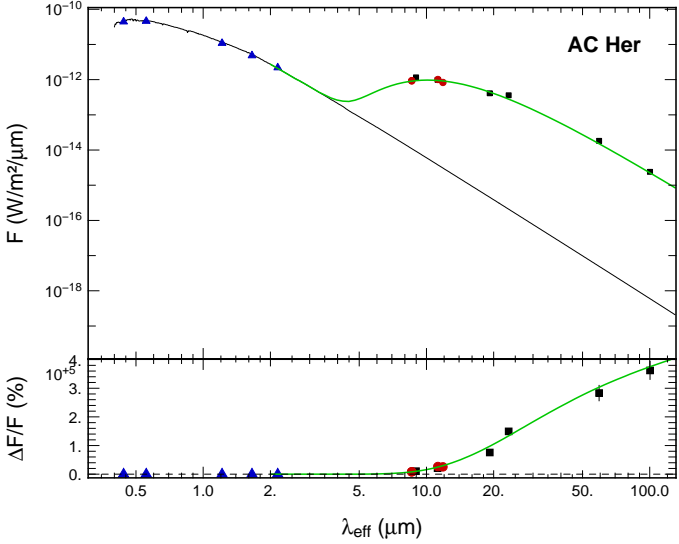}\hspace{.3cm}}
\caption{The same as Fig.~\ref{SED_R_SCT}, but for AC~Her.}
\label{SED_AC_HER}
\end{figure}

AC~Her has been studied extensively for its strong IR excess due to thermal emission from circumstellar dust grains \citep[see e.g.][]{de-Ruyter-2005-05}. We retrieved $B, V, J, H, K$ band photometric flux from \citet{Shenton-1992-08} (at phase 0.14). The longer wavelengths are data from \emph{IRC} (9 and 18\,$\mu$m), \emph{IRAS} (12, 25, 60, 100\,$\mu$m). E($B$-$V$) and $B$-$V$ were retrieve from \citet{Taranova-2010-02} and \citet{Myers-2001-09} respectively. We plotted the SED in Fig.~\ref{SED_AC_HER}.

As in \citet{de-Ruyter-2005-05} an infrared excess is clearly visible. Our values lie on the maximum of the spectral energy distribution and also show a strong contribution from the circumstellar environment (Table~\ref{measured_flux_densities}). We fitted a two component spectrum corresponding to the photosphere (still using Kurucz model) and the circumstellar emission. The best fitted parameters found for the photospheric emission (black solid curve in Fig.~\ref{SED_AC_HER}) with $\log{g} = 0.5$ from \citet{Van-Winckel-1998-08} are in Table~\ref{fitted_parameters}. The temperature is in agreement with the one estimated by \citet[][mean value: $T_\mathrm{eff} \sim 5400$\,K with $T_\odot = 5800$\,K]{Taranova-2010-02}. The angular diameter is consistent with \citet[][$\theta_\mathrm{LD} = 0.31$\,mas with $R = 23.6\,R_\odot$ and $d = 750\,\mathrm{pc}$]{Shenton-1992-08}. The second component was fitted with the model of Eq.~\ref{equation__blackbody}, assuming an optically thick medium (solid green curve in Fig.~\ref{SED_AC_HER}). The fitted temperature is listed in Table~\ref{fitted_parameters} with $A = 92.0 \pm 6.9\,\mathrm{mas^2}$. Our estimated temperature is larger than the one from \citet{Gielen-2007-11} who found a blackbody temperature of 170\,K using a disc model, his model also gave $R_\mathrm{in} = 50\,$mas and $R_\mathrm{out} = 428\,$mas for the disc (with $d = 1400\,$pc). However the author assumed that the morphology of circumstellar material is a disc. Other values are found from a dust shell model assumed by \citet{de-Ruyter-2005-05} who found $R_\mathrm{in} = 22\,$mas and $R_\mathrm{out} = 397\,$mas. \citet{Alcolea-1991-05} also fitted the extended emission with a shell model and found $R_\mathrm{in} = 76\,$mas and $R_\mathrm{out} = 3330\,$mas with a dust temperature of $T = 360\,$K. We can also notice that AC~Her was also studied using high-resolution imaging with adaptive optics by \citet{Close-2003-11} who did not detect any extended mid-infrared structure greater than 200\,mas.

\subsubsection{$\kappa$~Pav}

\begin{figure}[!h]
\centering
\resizebox{\hsize}{!}{\includegraphics{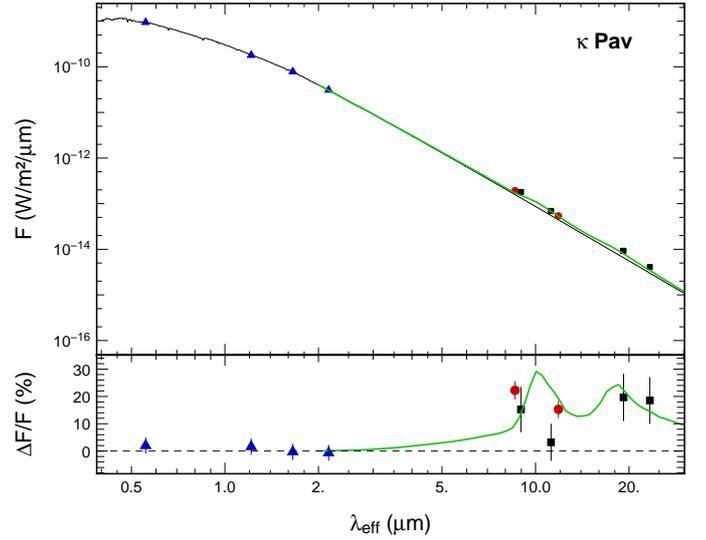}\hspace{.3cm}}
\caption{The same as Fig.~\ref{SED_R_SCT}, but for $\kappa$~Pav.}
\label{SED_K_PAV}
\end{figure}

\begin{figure*}[]
\centering
\resizebox{\hsize}{!}{\includegraphics{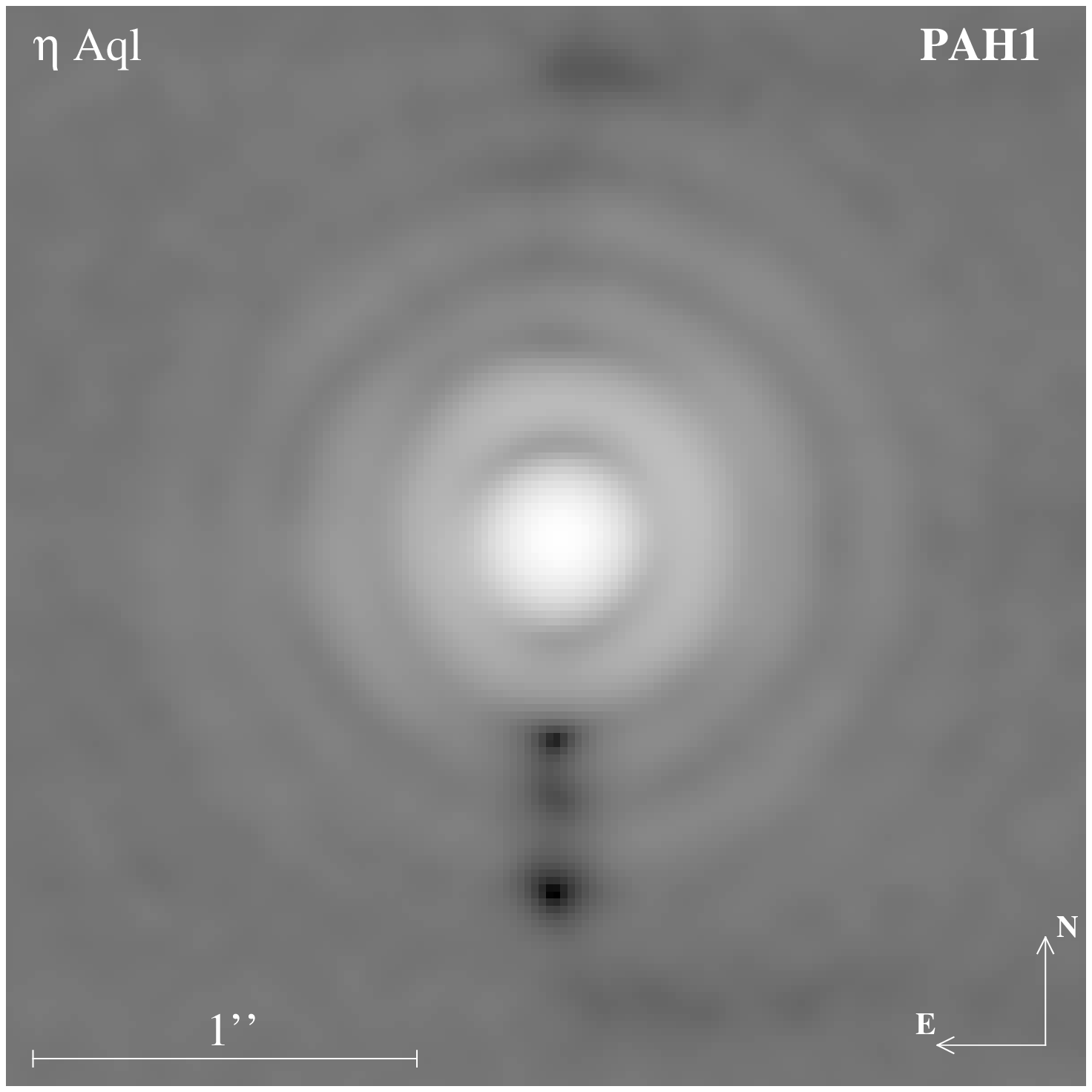}\hspace{.05cm}
\includegraphics{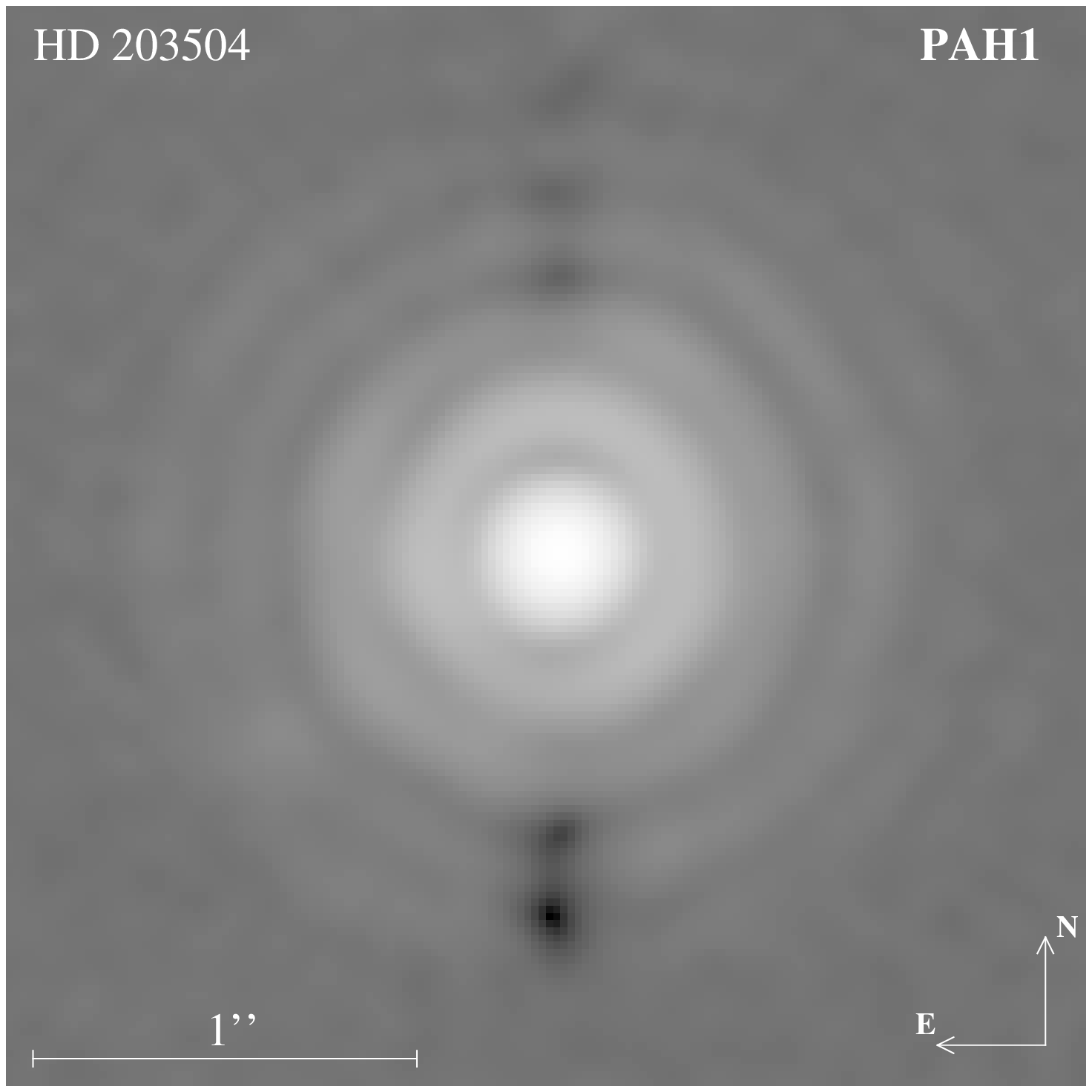}\vspace{.05cm}
\includegraphics{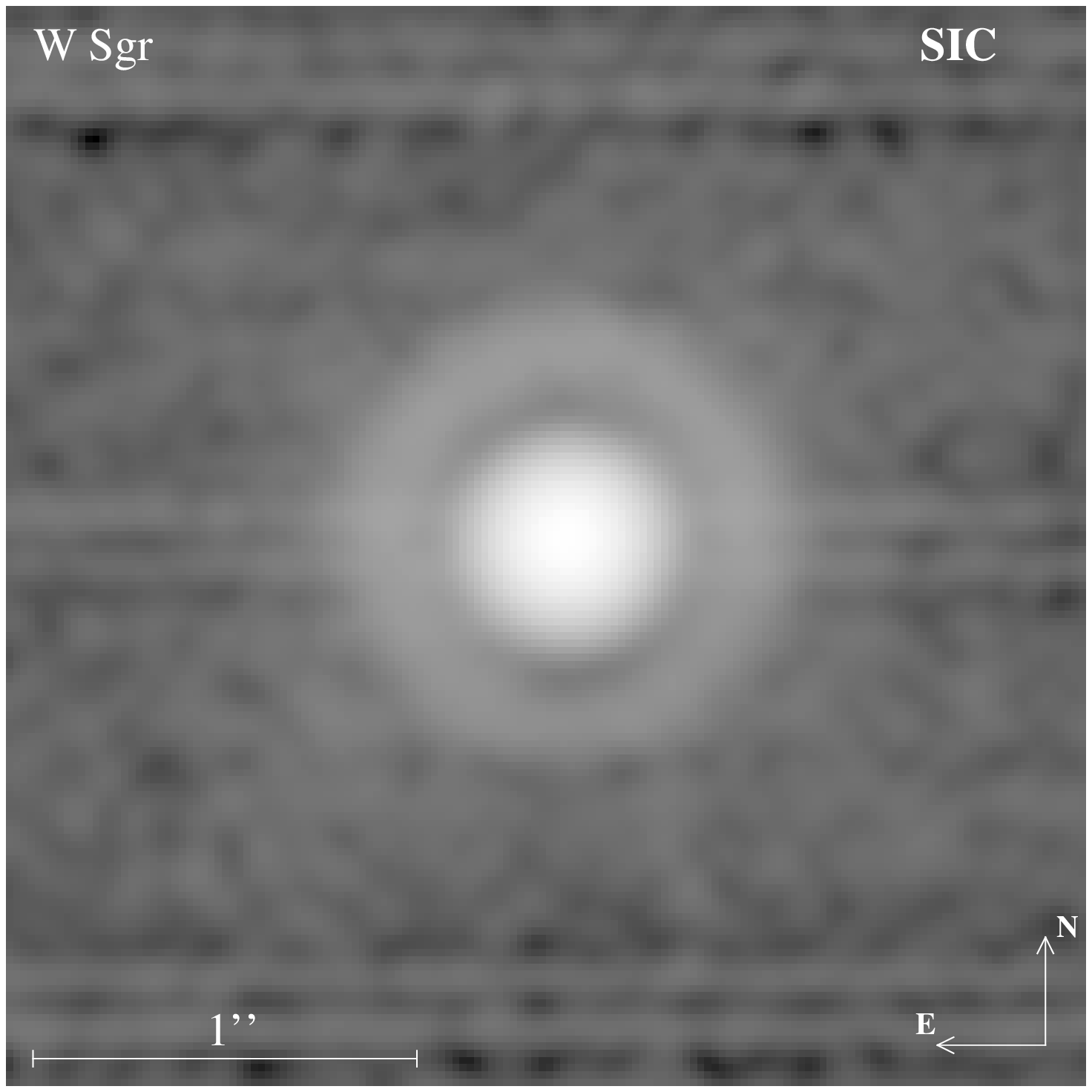}\vspace{.05cm}
\includegraphics{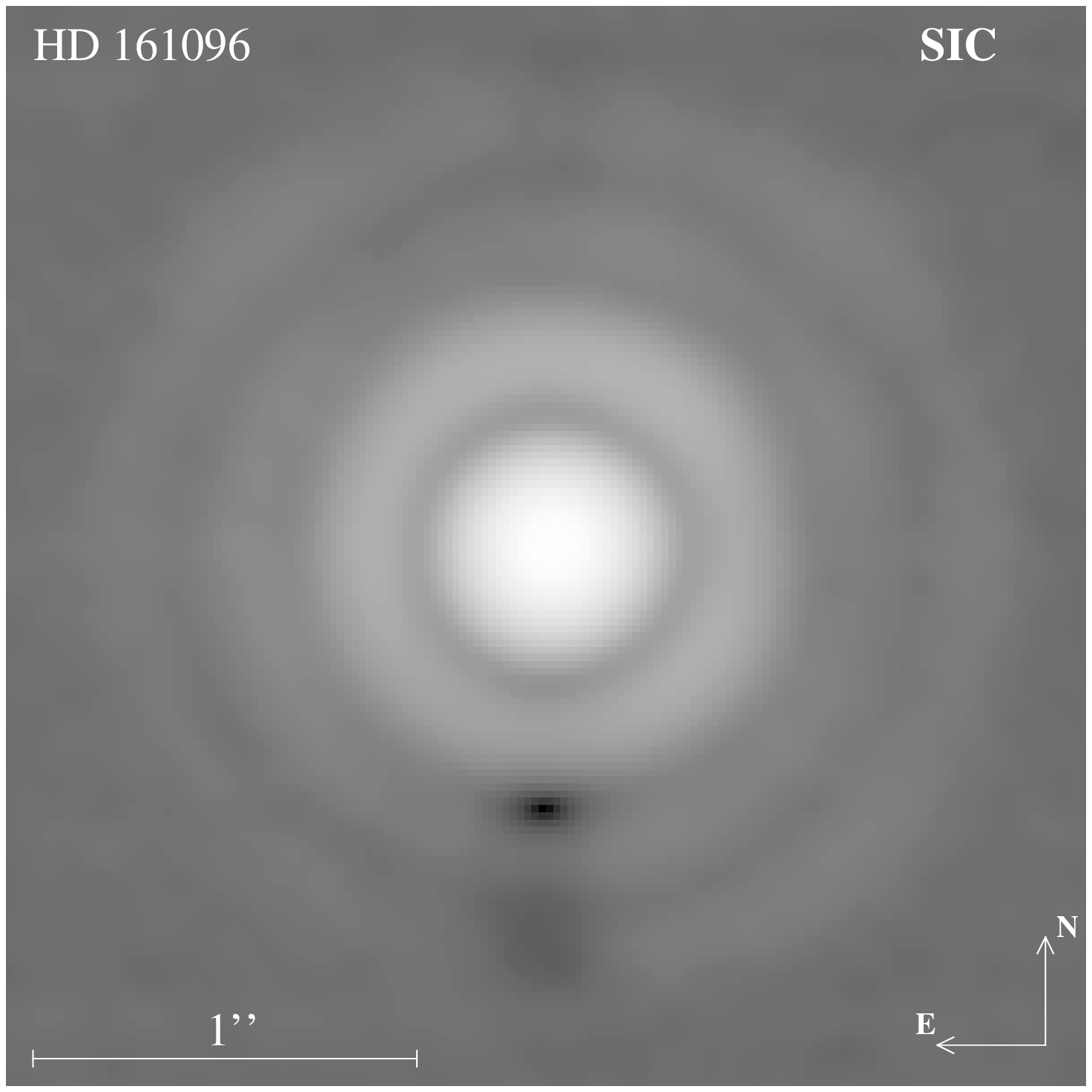}\vspace{.05cm}}
\resizebox{\hsize}{!}{\includegraphics{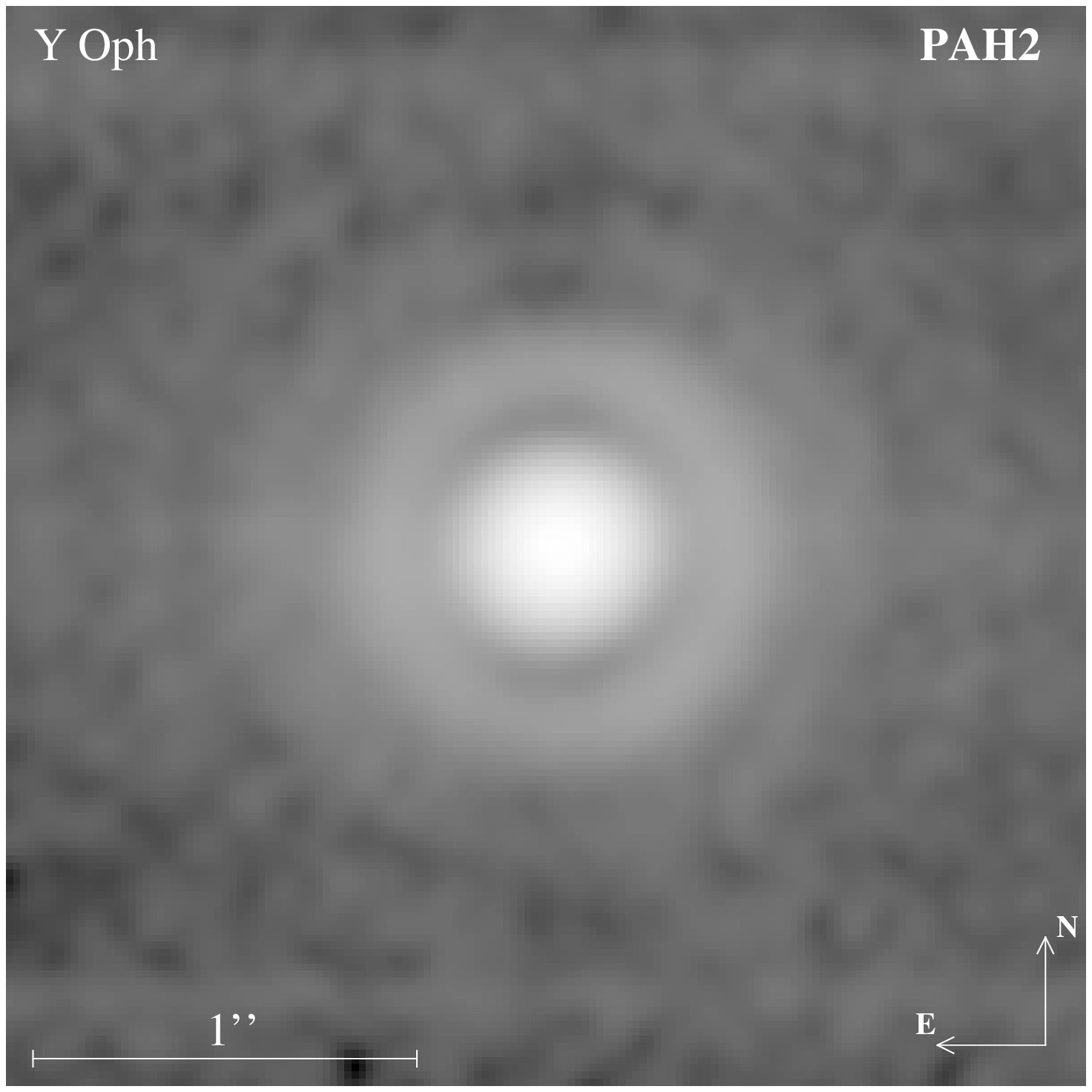}\hspace{.05cm}
\includegraphics{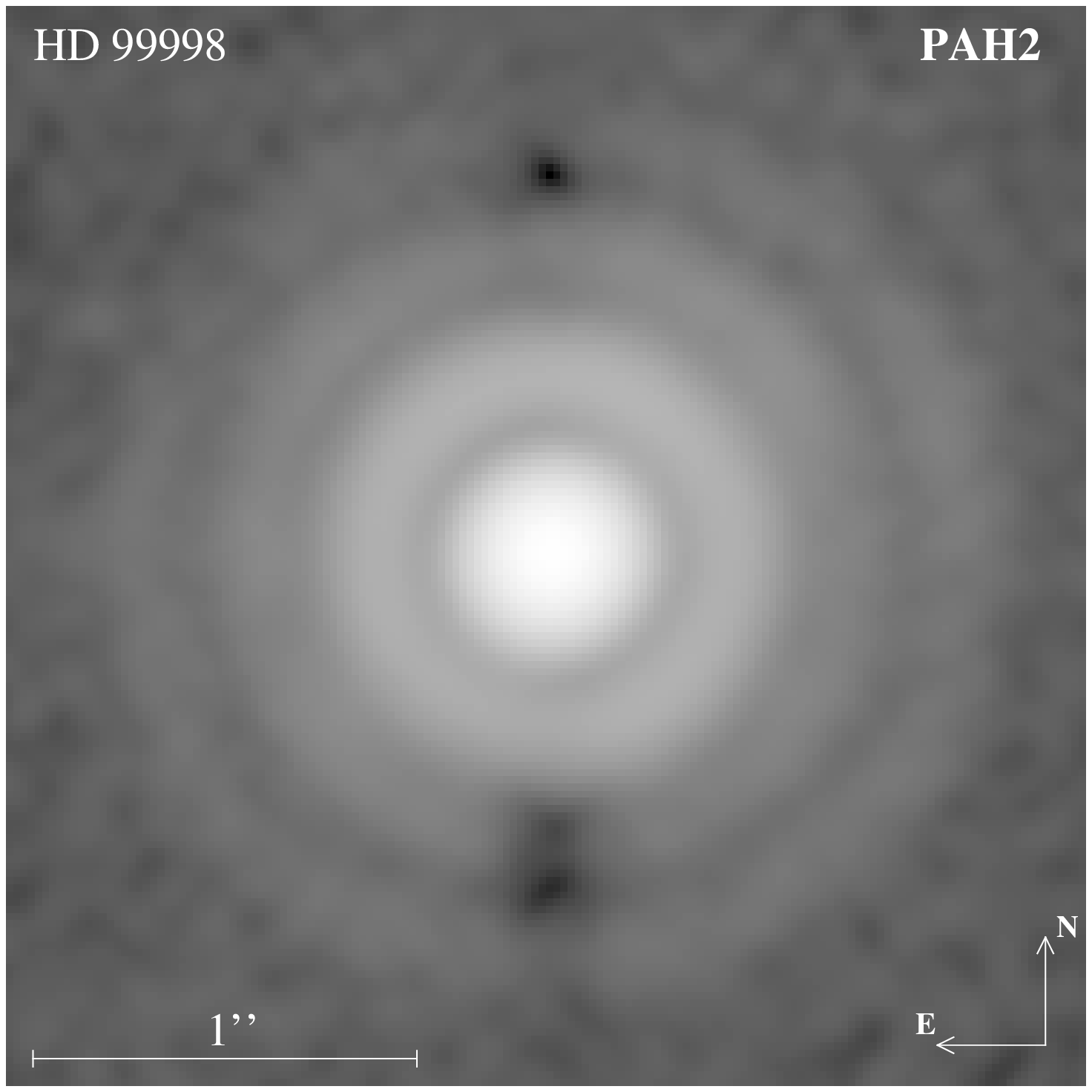}\hspace{.05cm}
\includegraphics{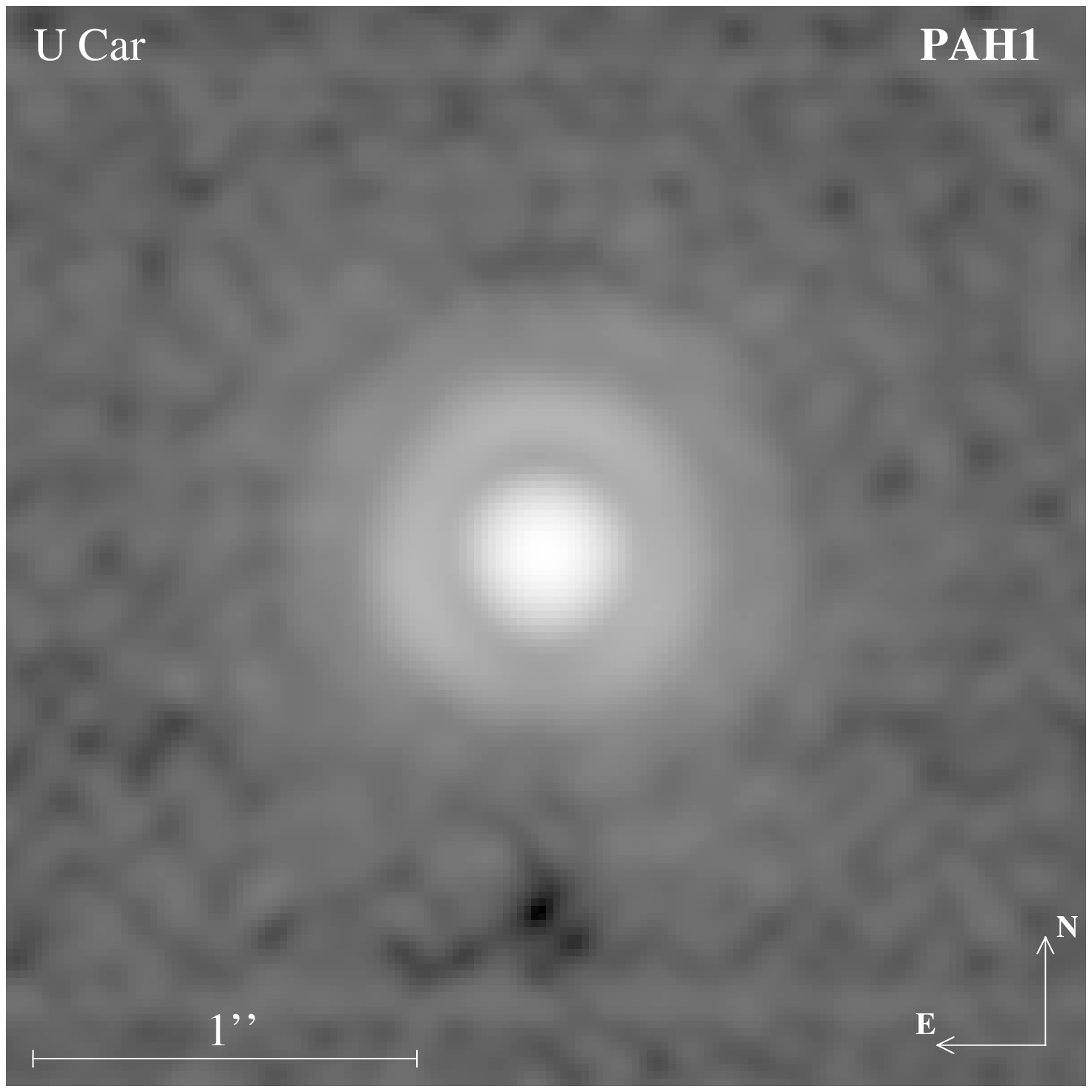}\vspace{.05cm}
\includegraphics{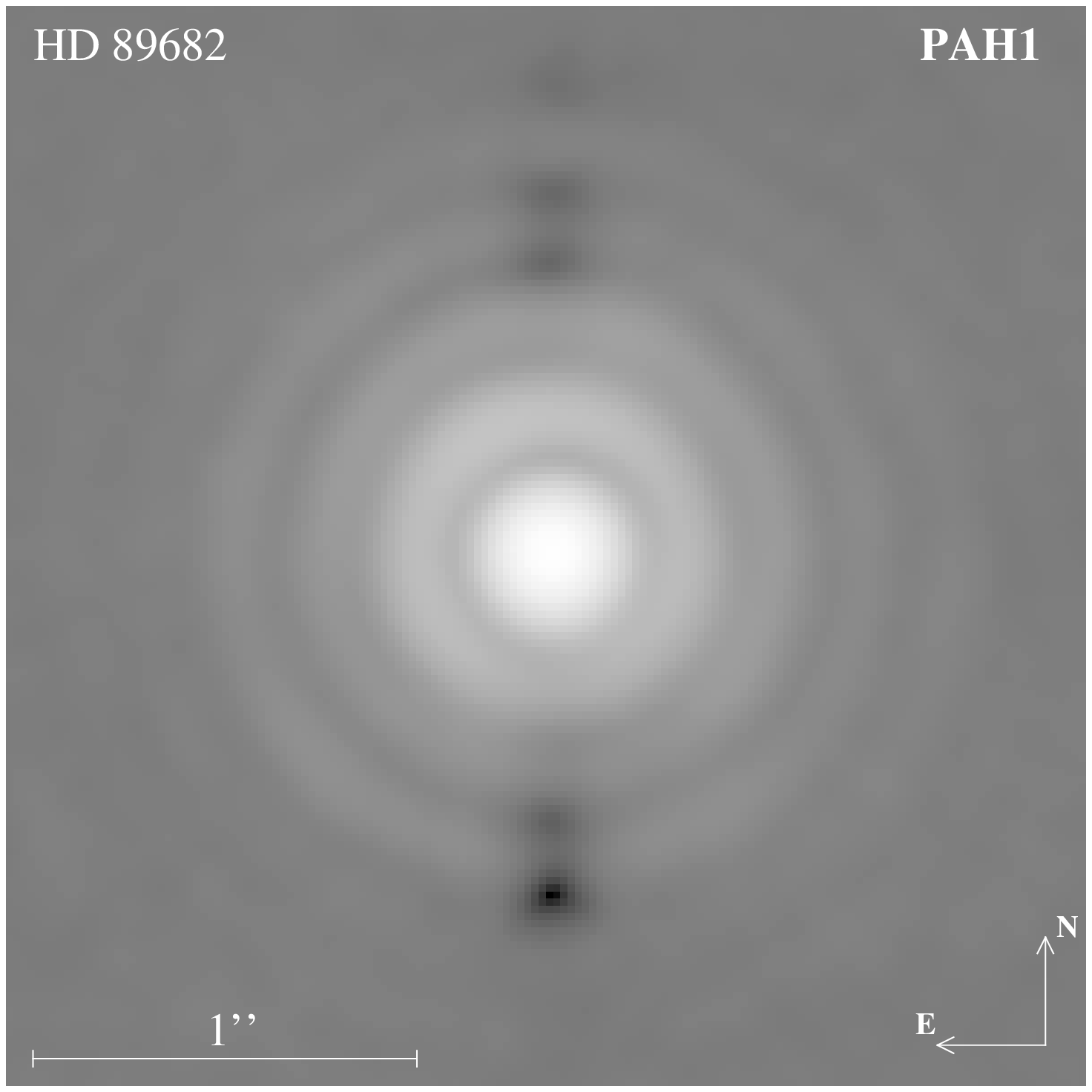}\vspace{.05cm}}
\resizebox{\hsize}{!}{\includegraphics{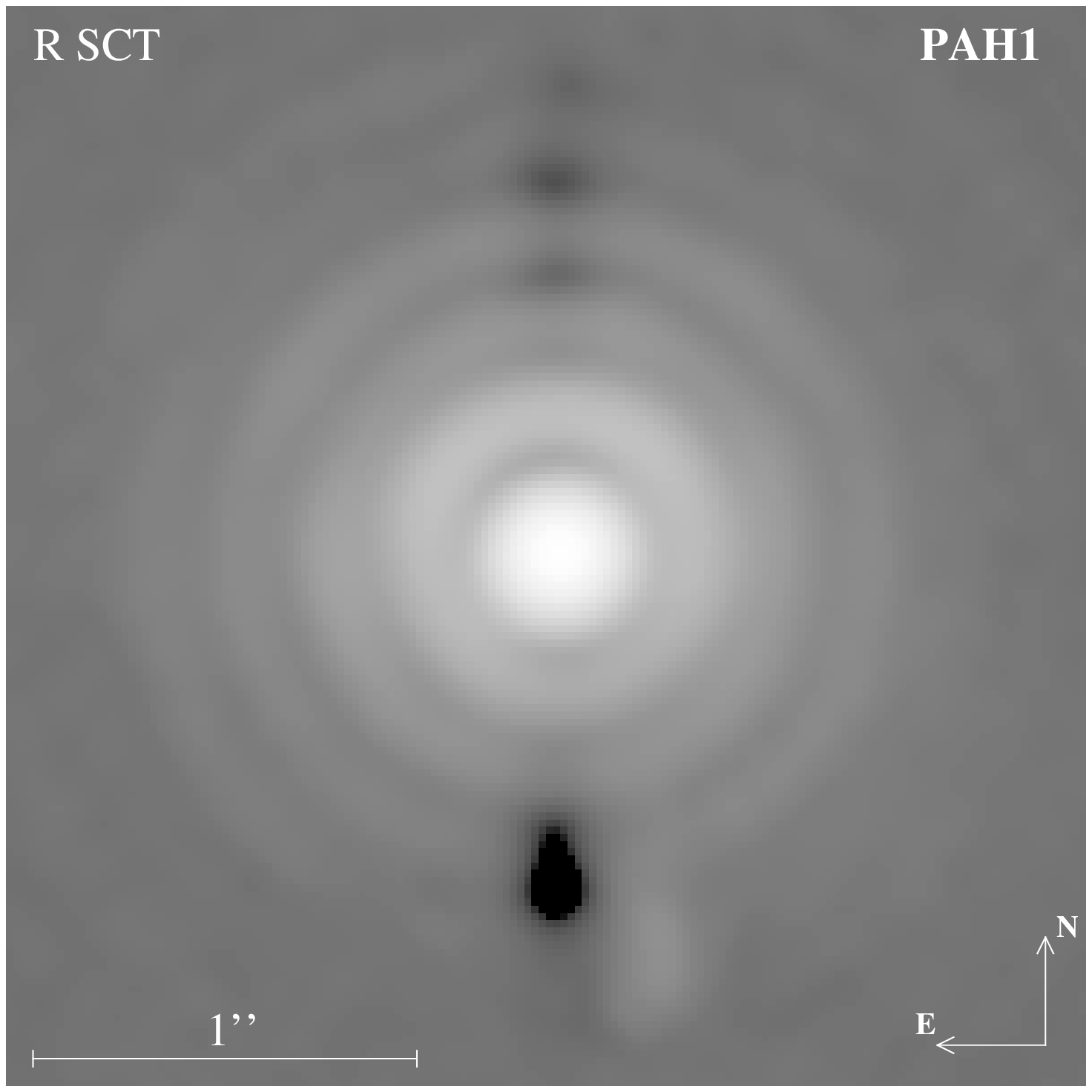}\hspace{.05cm}
\includegraphics{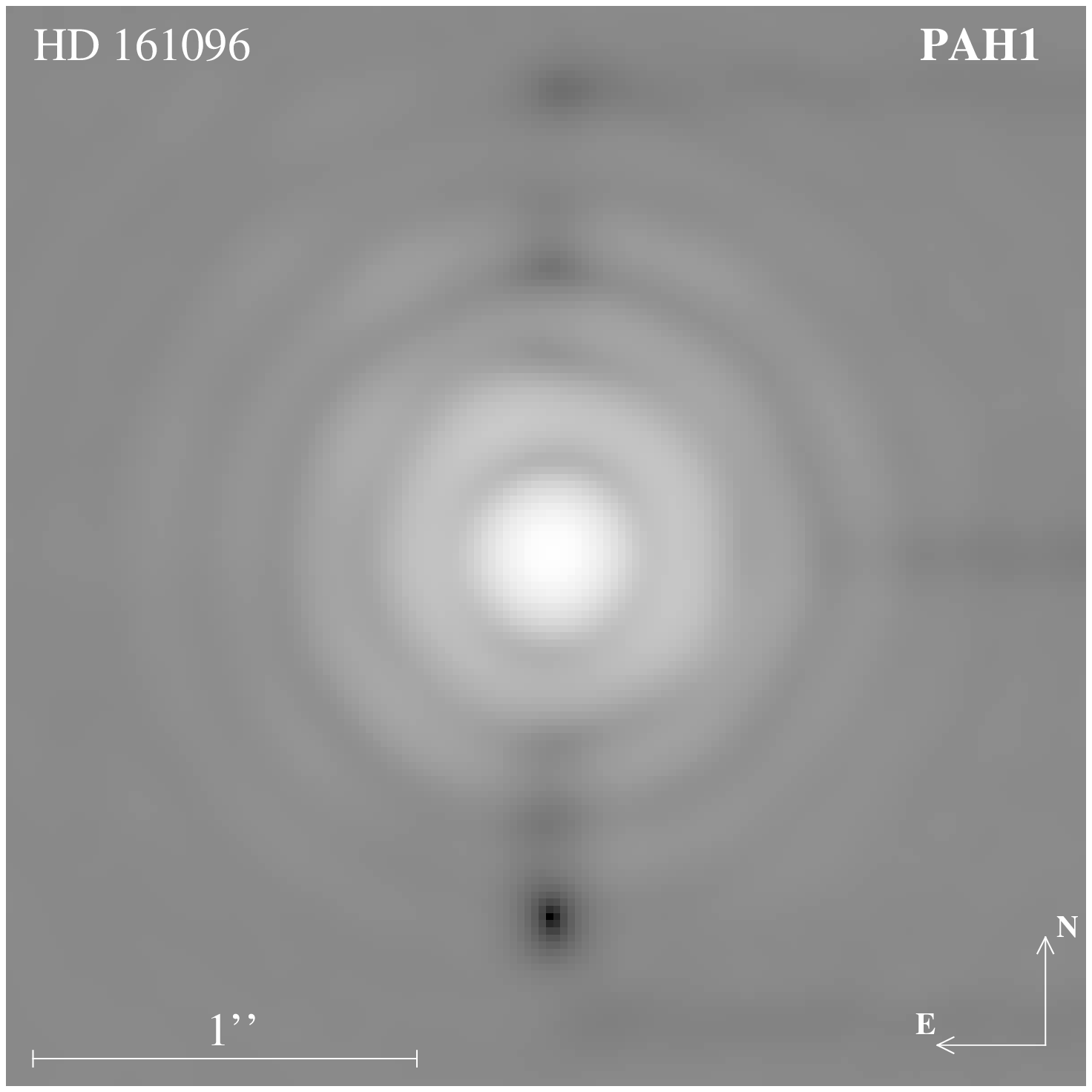}\hspace{.05cm}
\includegraphics{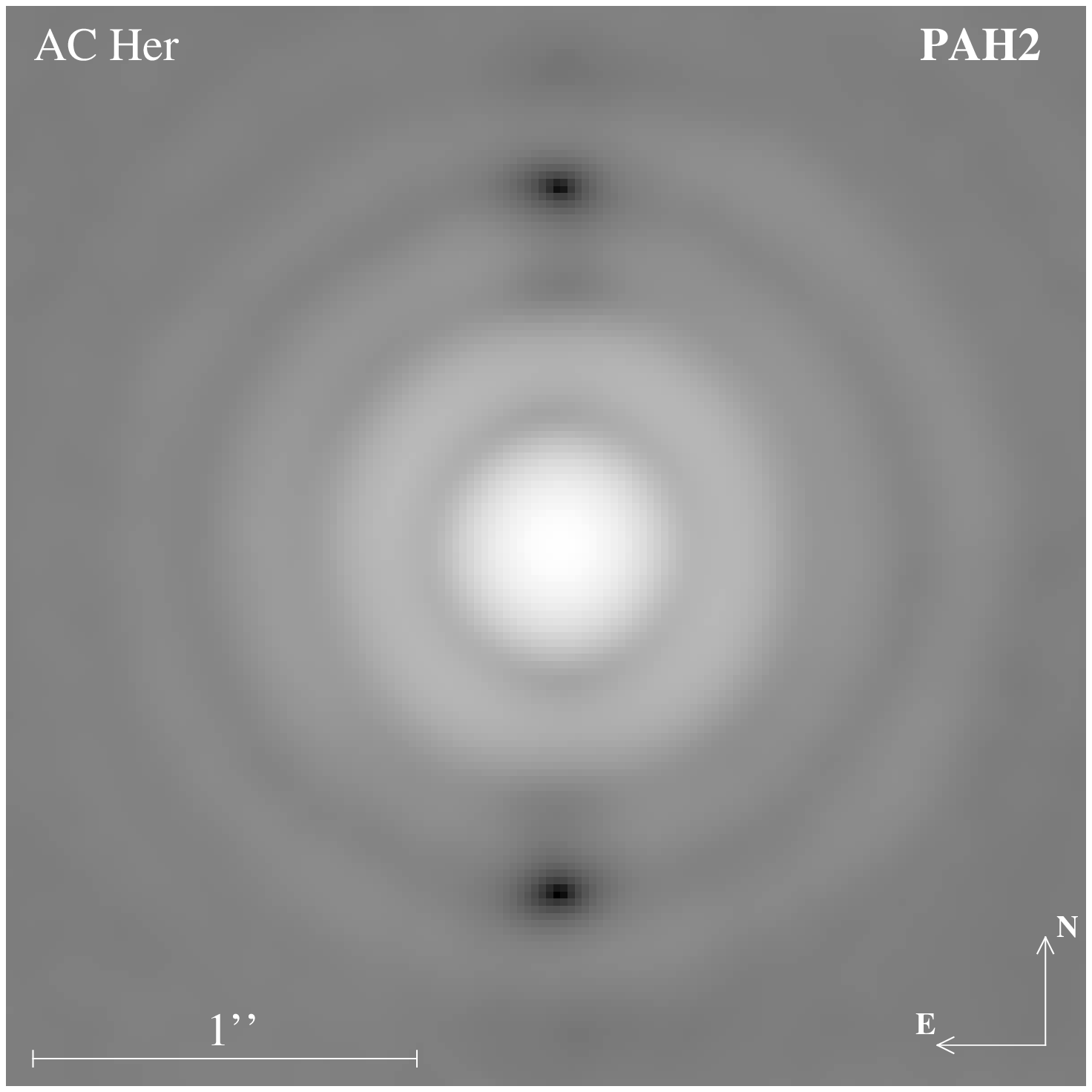}\hspace{.05cm}
\includegraphics{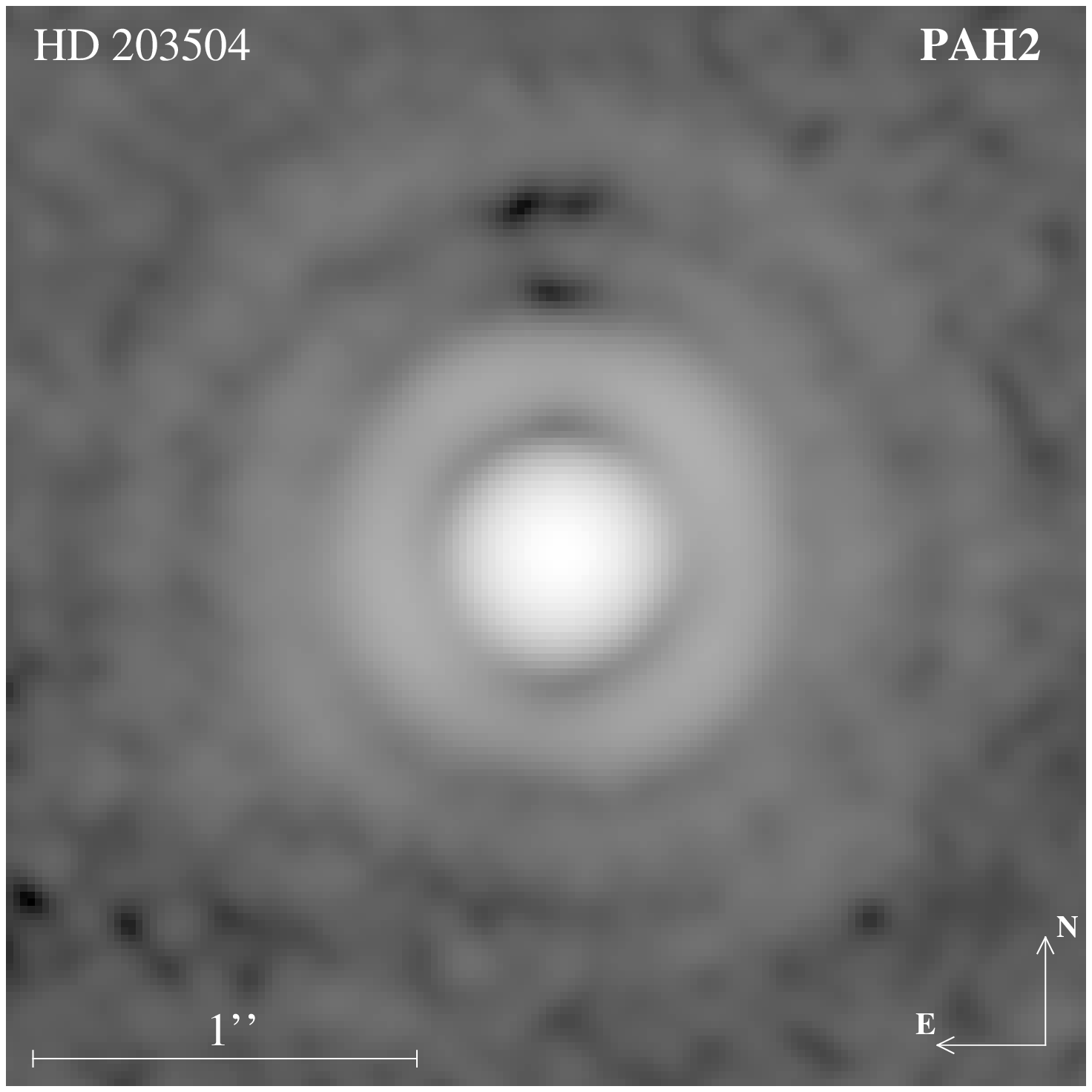}\hspace{.05cm}}
\caption{Sample of the final VISIR images for our Cepheids and their calibrator. The scale is logarithmic for all images.}
\label{VISIR_images}
\end{figure*}

We retrieved the $V, J, H, Ks$ photometric light curves from \citetalias{Feast-2008-06-2}. We used E$(B-V)$ end $B-V$ from \citetalias{Feast-2008-06-2}. The mid- and far-infrared data were obtained from \emph{IRC} and \emph{IRAS}. We plotted the SED in Fig.~\ref{SED_K_PAV}.

We chose a mean effective gravity of $\log{g} = 1.2$ from \citet{Luck-1989-07}. Our fitted parameters are presented in Table~\ref{fitted_parameters}. Our diameter estimate is in good agreement with \citetalias{Feast-2008-06-2} ($1.04 \pm 0.04$\,mas at $\phi = 0.90$, with $d = 204 \pm 6$\,pc). However we found an effective temperature 9\,\% larger than the mean value estimated by \citet[][$T_\mathrm{eff} \sim 5750$\,K at $\phi = 0.94$]{Luck-1989-07} based on metal abundance analyses. It has to be noted that this star is classified as peculiar type II Cepheid with distinctive light curve and brighter than normal type II Cepheid with the same period \citep[see e.g.][]{Matsunaga-2009-08}.

We detected with \emph{VISIR} an IR excess of the order of 20\,\%. This excess tends to increase with wavelength leading to the hypothesis of an extended circumstellar envelope. We fitted to the IR photometric data the model of Eq.~\ref{equation__greybody}, assuming the envelope is optically thin. As some features are present around $10\,\mu\mathrm{m}$ and $20\,\mu\mathrm{m}$, we also assumed a composition of silicate/carbon grains. We used the dust opacity of \citetalias{Ossenkopf-1994-11} to derive the total dust mass, assuming a gas to dust ratio of 100. The best fitted parameters are listed in Table~\ref{fitted_parameters}. 


\subsection{Spatially resolved emission}
\label{spatially_resolved_emission}

We search for spatially extended emission using a Fourier technique, similar in its principle to the calibration technique used in long baseline interferometry. This method was already used and validated by \citet{Kervella-2009-05,Kervella-2006-07}. The principle is to divide the Fourier transform modulus of the image of the Cepheid ($I_\mathrm{cep}$) by that of the calibrator image ($I_\mathrm{cal}$):

\begin{displaymath}
\Psi (\nu_x,\nu_y) = \left| \frac{ \hat{I}_{\mathrm{cep}}(x,y) }{ \hat{I}_{\mathrm{cal}}(x,y) } \right|
\end{displaymath}
where the hat symbol denote the Fourier transform, $(x, y)$ the sky coordinates and $(\nu_x,\nu_y)$ the angular spatial frequencies. This equation is related to interferometric observations that provide measurements of the Fourier transform of the intensity distribution of the observed object (Van Cittert-Zernike theorem).

We then compute the ring median of $\Psi$, i.e. the median for a given spatial frequency radius $\nu$ over all azimuths (where $\nu^2 = \nu_x^2 + \nu_y^2$). The function $\Psi(\nu)$ obtained is equivalent to a visibility in interferometry. The error bars on $\Psi$ were estimated by the quadratic sum of the dispersion of the PSF calibrator’s Fourier modulus over the night and the rms dispersion of the calibrated $\Psi$ function over all azimuth directions for each spatial frequency. A deviation from a central symmetry will not be detected and any departure will be included in the error bars.

Defining a model of a point-like star surrounded by a Gaussian shaped CSE and taking its Fourier transform, it is possible to retrieve the CSE intensity distribution. This type of model was already used by \citet{Kervella-2009-05} and \citet{Kervella-2006-07} with the $\Psi(\nu)$ function

\begin{displaymath}
\Psi (\nu,\rho_\lambda,\alpha_\lambda) = \frac{f_\star V_\star + f_\mathrm{cse}V_\mathrm{cse}}{f_\star + f_\mathrm{cse}}
\end{displaymath}
\begin{displaymath}
\Psi (\nu,\rho_\lambda,\alpha_\lambda) = \frac{1}{1 + \alpha_\lambda} \left[ 1 + \alpha_\lambda \exp{\left(-\frac{(\pi\ \rho_\lambda\ \nu)^2}{4\ln{2}} \right) }\right]
\end{displaymath}
where the Gaussian CSE is defined with a FWHM $\rho_\lambda$ and a relative flux $\alpha_\lambda = f_\mathrm{cse}(\lambda)/f_\star(\lambda)$, i.e the ratio of the flux of the envelope to the photospheric flux. We set $V_\star = 1$ since the star is unresolved by \emph{VISIR}.

\begin{figure*}[!ht]
\centering
\includegraphics[width=7.5cm]{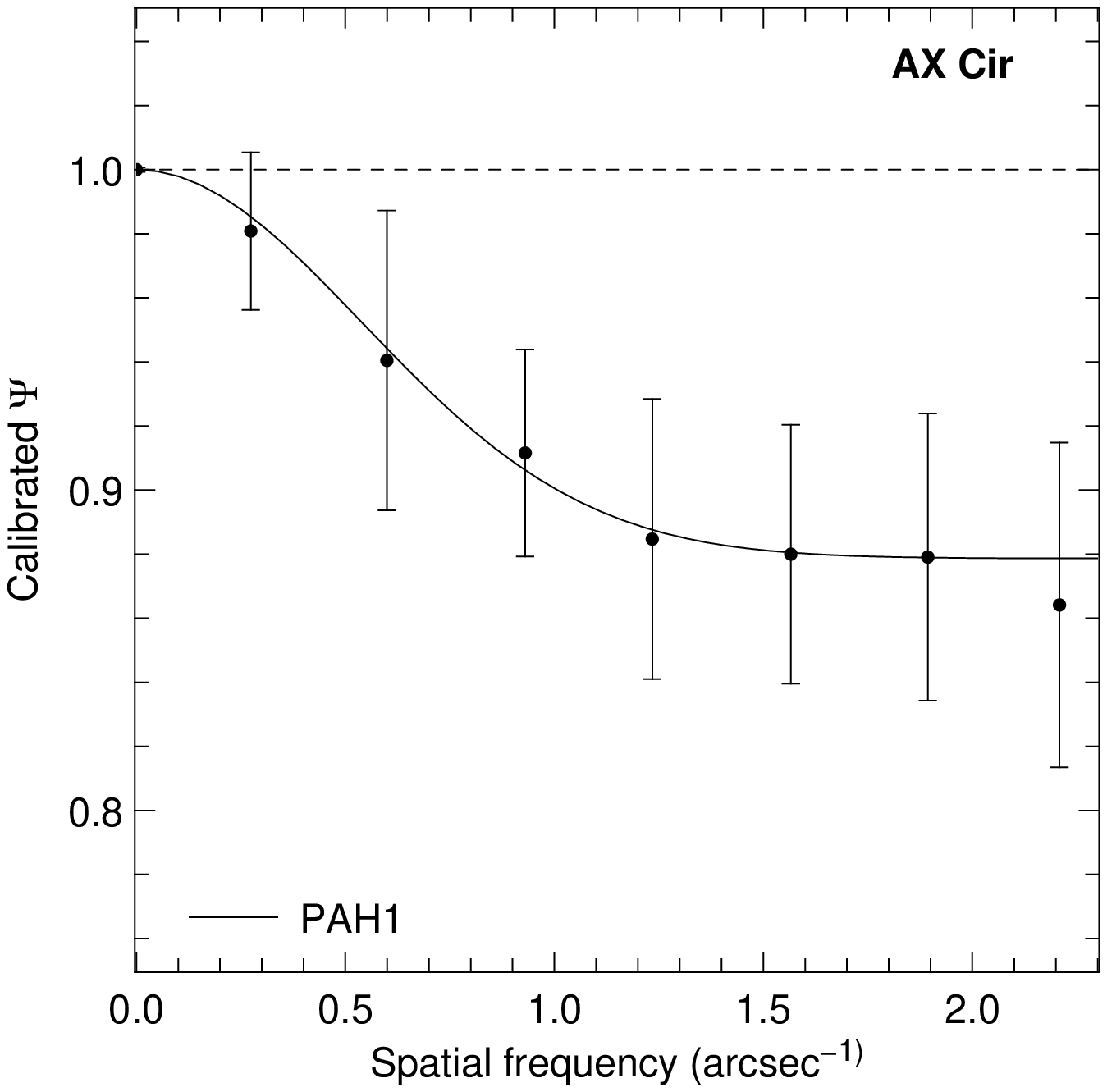}\hspace{.5cm}
\includegraphics[width=7.5cm]{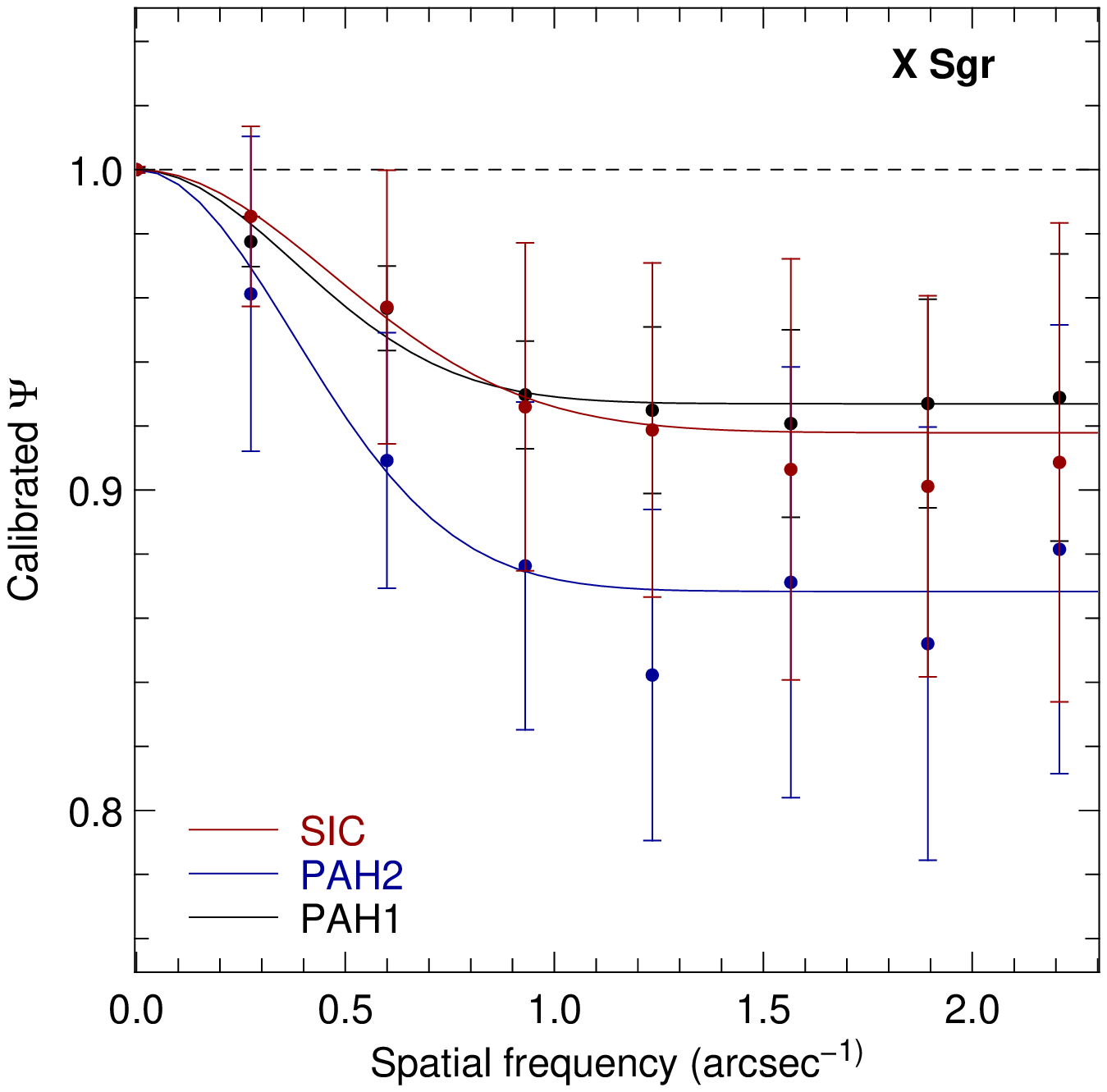}\vspace{.5cm}
\includegraphics[width=7.5cm]{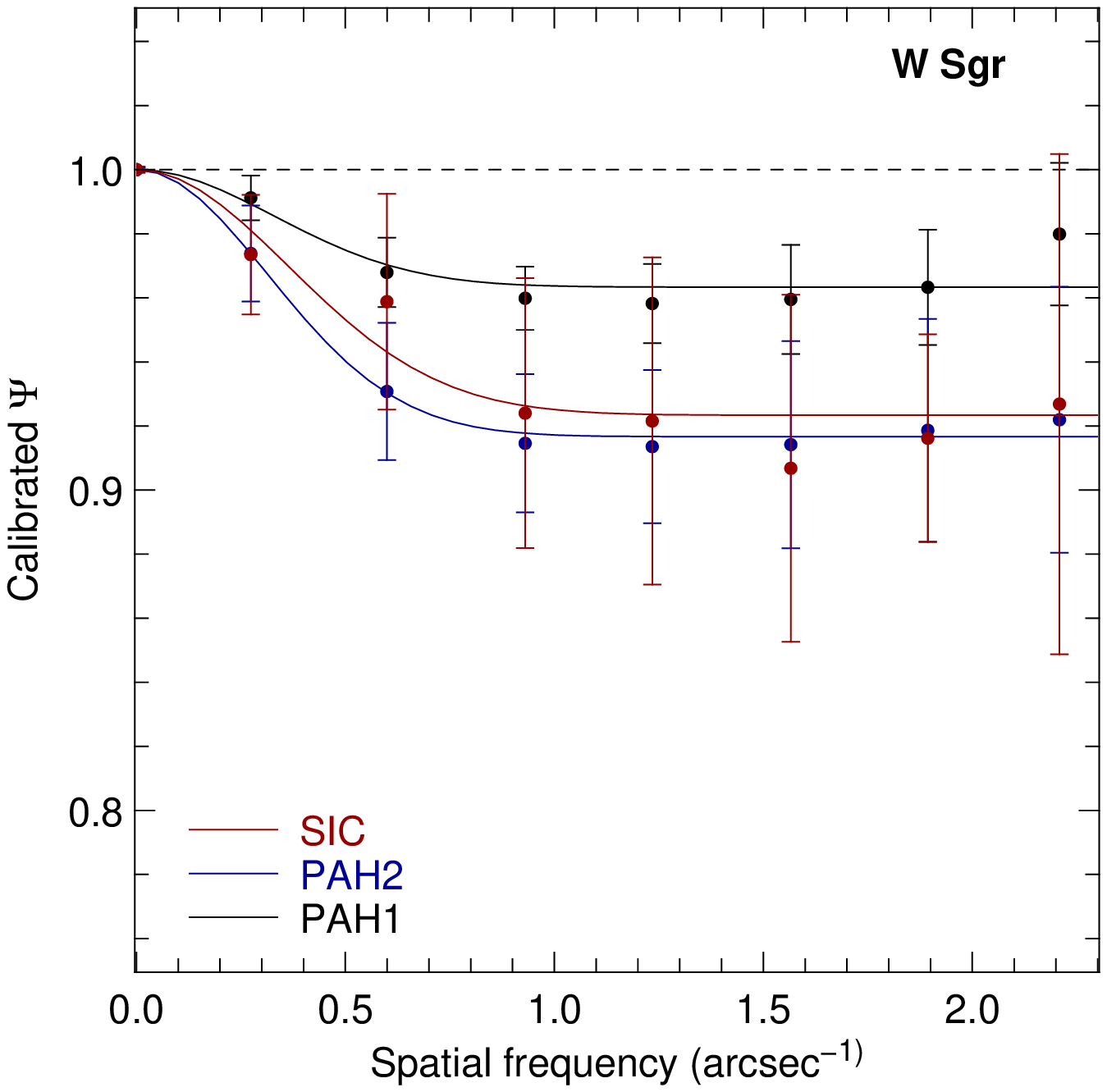}\hspace{.5cm}
\includegraphics[width=7.5cm]{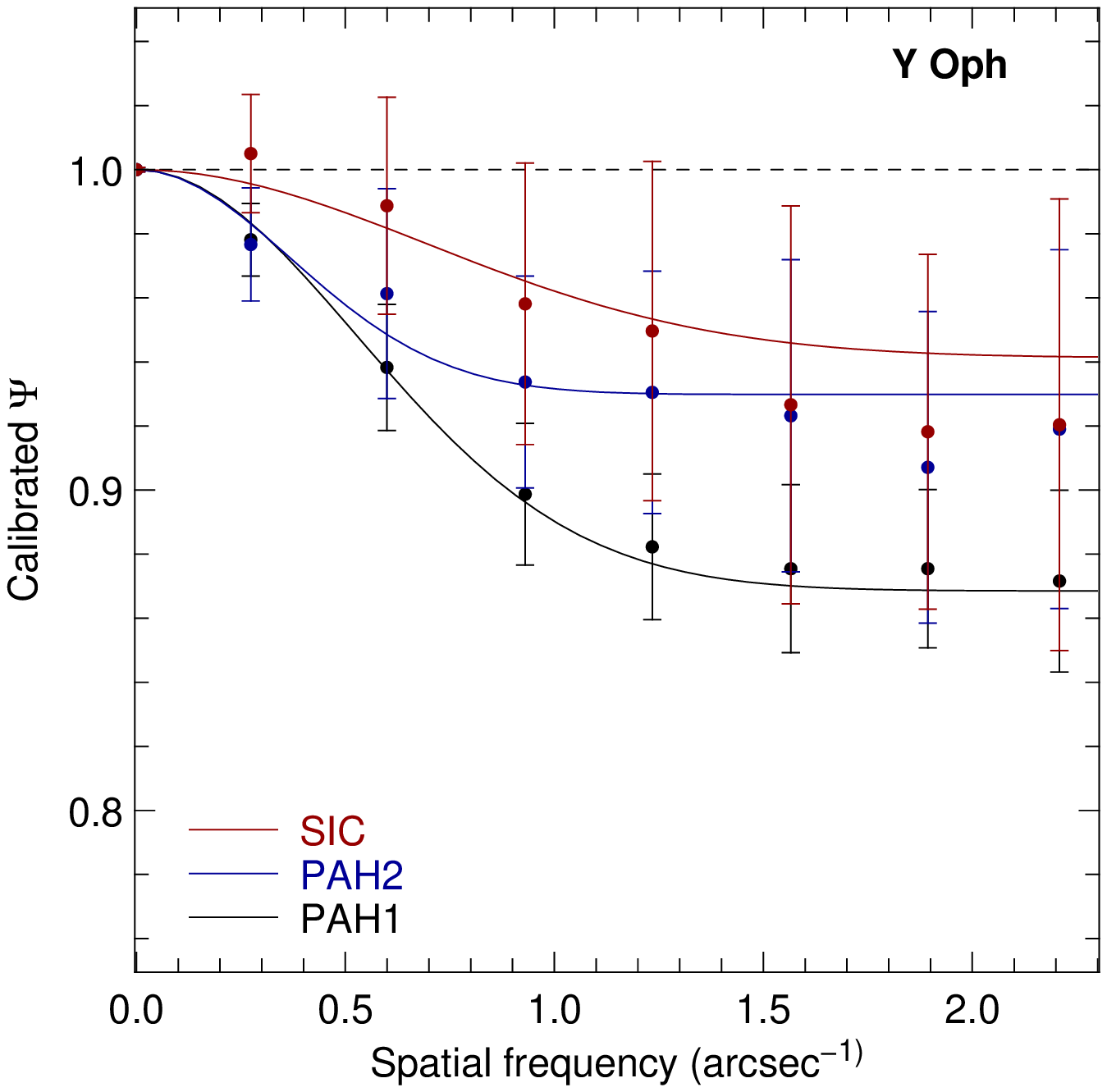}\vspace{.5cm}
\includegraphics[width=7.5cm]{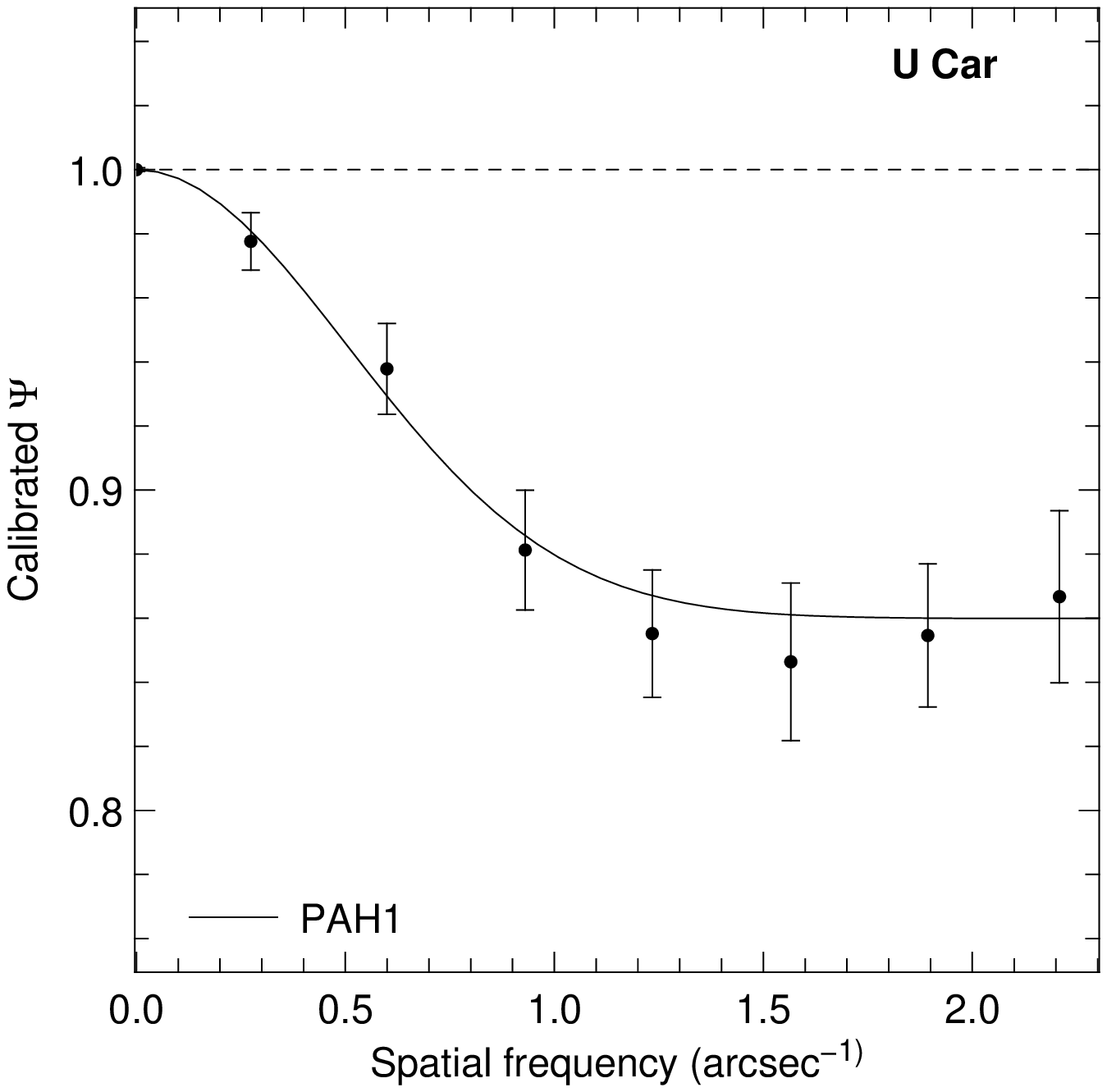}
\caption{Fit of a Gaussian CSE + unresolved point source model to the $\Psi$ functions of our Cepheids. The dashed curve is here for a reference as an unresolved star.}
\label{psi_function}
\end{figure*}

We applied the fit using a classical $\chi^2$ minimization to all the final images. We did not detect spatially resolved emission for FF~Aql, $\eta$~Aql, U~Car (in the SiC filter), SV~Vul, R~Sct, AC~Her and $\kappa$~Pav. For the other Cepheids that show a resolved component, the fitted parameters are presented in Table~\ref{gaussian_parameter} and the $\Psi$ function are plotted in Fig.~\ref{psi_function}.

\begin{table}[!h]
\centering
\caption{Fitted parameters of the $\Psi (\nu,\rho_\lambda,\alpha_\lambda)$ function.}
\begin{tabular}{cccccc} 
\hline
\hline
Stars	  					&	Filter			& $\rho$						& $\alpha$ 					\\
 		  					&					& 	 (\arcsec)					& (\%)							\\
\hline
AX~Cir					&  PAH1		&	$0.69 \pm 0.24$		&	$13.8 \pm 2.5$			\\
\hline
X~Sgr					&  PAH1		&	$0.99 \pm 0.23$		&	$7.9 \pm 1.4$			\\
							&  PAH2		&	$0.99 \pm 0.44$		&	$15.2 \pm 3.7$			\\
							&  SiC			&	$0.81 \pm 0.53$		&	$8.9 \pm 3.5$			\\
\hline
W~Sgr					&  PAH1		&	$1.14 \pm 0.39$		&	$3.8 \pm 0.6$			\\
							&  PAH2		&	$1.19 \pm 0.37$		&	$9.1 \pm 1.5$			\\
							&  SiC			&	$1.03 \pm 0.50$		&	$8.3 \pm 2.4$			\\
							\hline
Y~Oph					&  PAH1		&	$0.71 \pm 0.12$		&	$15.1 \pm 1.4$			\\
							&  PAH2		&	$1.02 \pm 0.52$		&	$7.5 \pm 2.3$			\\
							&  SiC			&	$0.54 \pm 0.46$		&	$6.2 \pm 4.1$			\\
\hline	
U~Car					&  PAH1		&	$0.74 \pm 0.10$		&	$16.3 \pm 1.4$			\\
\hline
\end{tabular}
\label{gaussian_parameter}
\end{table}


For the stars with an undetected envelope, an upper limit of $\sim 265$\,mas can be set for the extension of the CSE based on the telescope resolution.

The fact that we detect an extended emission around AX~Cir and not in the previous section (Sect.~\ref{spectral_energy_distribution}) can be explain by the lack of $J, H, K$ light curves to fully constrain the SED at our phase of pulsation. The uncertainties in these bands (0.14\,mag in $K$) are large enough to prevent a detection of a IR excess.

From the distance $\rho$ estimated with this technique, it is possible to assess a lower limit for the mass loss rate by computing the time $t$ required by a stellar wind to reach this distance. This computation has already been applied by \citet{Marengo-2010-12} for the Cepheid $\delta$~Cep. Using the escape velocity $v_\mathrm{esc} \sim 100\,\mathrm{km\,s^{-1}}$ \citep{Welch-1988-06} as the minimum wind speed, we found $t \sim 12$--$23$\,yr. With the total dust mass estimated in Sect.~\ref{spectral_energy_distribution}, we derive a minimum mass loss rate $\dot{M} \sim 7\times10^{-11}\,M_\odot\,yr^{-1}$ for AX~Cir and W~Sgr, and $\dot{M} \sim 2\times10^{-10}\,M_\odot\,yr^{-1}$ for X~Sgr, Y~Oph and U~Car. These values are comparable with the predicted mass loss rate from \citet[][ranging from $10^{-10}$ to $10^{-7}\,M_\odot\,yr^{-1}$]{Neilson-2008-09} and the measured mass loss rate from \citet[][ranging from $10^{-10}$ to $10^{-6}\,M_\odot\,yr^{-1}$]{Deasy-1988-04}.

The Fourier technique gives IR excesses that are very comparable to those derived from the SED fitting presented in Sect.~\ref{spectral_energy_distribution} (except for AX Cir). This convergence of two independent analysis methods gives us good confidence that the derived envelope parameters are reliable.

\section{Discussion}
\label{discussion}

Classical and type II Cepheids are two distinct classes of pulsating stars. The formers are know to be intermediate mass stars with regular pulsation periods, evolving as post-main sequence stars in a core He-burning phase. Type II Cepheids are generally associated to lower mass stars with irregular pulsations and undergoing the post-core He-burning evolution. Therefore the CSEs surrounding these two classes correspond to different evolutionary stages.

R~Sct and AC~Her have been observed extensively and are known to have strong IR excesses linked to the extended CSEs which are interpreted as being a relic of their strong dusty mass loss on the asymptotic giant branch. Different models have been applied to constrain the morphology of this surrounding material. \citet{de-Ruyter-2005-05} interpreted the spectral energy distribution of these stars using a dust shell model. They found an inner radius $R_\mathrm{in} = 12.5$\,AU and an outer radius $R_\mathrm{out} = 224$\,AU for AC~Her and $R_\mathrm{in} = 13.4$\,AU and $R_\mathrm{out} = 5200$\,AU for R~Sct. From a disc model applied to AC~Her, \citet{Gielen-2007-11} found $R_\mathrm{in} = 35$\,AU and a smaller external radius $R_\mathrm{out} = 300$\,AU. For this star these results are in contradiction with \citet{Close-2003-11} who excluded any extended emission larger than 75\,AU from high Strehl ratio adaptive optics images in the mid-IR domain. We also not detected in our filters with \emph{VISIR} extended emission larger than $R = 100$\,AU.

The case of R~Sct is peculiar and this star is often classified as an exception by several authors. \citet{Alcolea-1991-05} used a two shell model to interpret both the mid-IR and far-IR data and they estimated a temperature for the largest grains of $T = 815\,$K. The estimated radii of the shells were $R_\mathrm{in,1} = 30$\,AU, $R_\mathrm{out,1} = R_\mathrm{in,2} = 5\,720$\,AU and $R_\mathrm{out,2} = 12\,000$\,AU (using $d = 431\,$pc) where indexes 1 and 2 denote the inner and outer shell respectively. We possibly detected the emission of the inner shell from our \emph{VISIR} photometry but no spatially extended emission larger than $R = 100$\,AU has been detected.

$\kappa$~Pav has been recently classified as a peculiar type II Cepheid by some authors because of its distinctive light curve. It is significantly brighter than normal type II Cepheids with the same period \citep[see e.g.][]{Matsunaga-2009-08}. This star is not at the same stage of evolution as R~Sct and AC~Her since it is ascending in the Hertzsprung–Russell diagram the blue horizontal branch to the asymptotic giant branch. During this process the star undergoes changes both in its core and envelope that could lead to mass loss via pulsation and/or shock mechanisms. An active past and/or on-going mass loss may explain the large IR excess we have detected.

With classical Cepheids we are on different scales since the envelope that have been detected up to now only have a spatial extension of a few stellar radius. Only a small sample of classical Cepheids is known to host a circumstellar envelopes \citep[\object{$\ell$~Car}, Y~Oph, \object{RS~Pup}, \object{Polaris}, \object{$\delta$~Cep}, \object{S~Mus}, \object{GH~Lup}, \object{T~Mon} and \object{X~Cyg},][]{Kervella-2006-03,Merand-2006-07,Merand-2007-08,Barmby-2010-11}. With this paper we extend this sample with FF~Aql, $\eta$~Aql, W~Sgr, U~Car and SV~Vul as probable IR excess Cepheids. The origin of these CSEs is not well understood. 
Their presence could be linked to mass loss from the star. It has been suggested that the IR excess is caused by dust formed in a wind from the Cepheids \citep{Kervella-2006-03}. From a radiative-driven wind model including pulsation and shock effects \citet{Neilson-2008-09} concluded that radiative driving is not sufficient to account for the observed IR excesses and proposed that the mass loss could be driven by shocks generated in the atmosphere by the pulsation of the star. Other observational evidences were provided with IR excess detections from \emph{IRAS} observations by \citet{Deasy-1988-04} who estimated mass loss rates ranging from $10^{-10}$ to $10^{-6}$\,$M_\odot$\ yr$^{-1}$.

\begin{figure}[!h]
\resizebox{\hsize}{!}{\includegraphics[width=8.5cm]{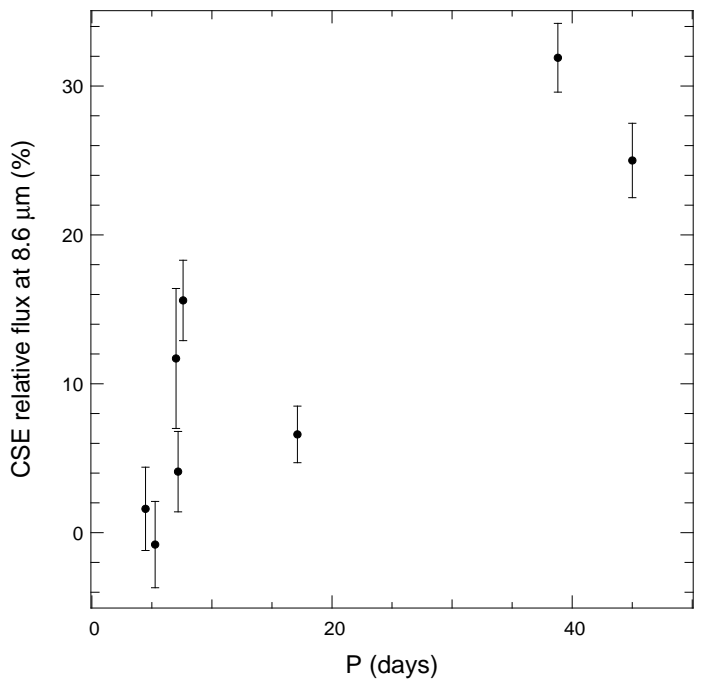}}
\caption{Measured relative CSE fluxes at 8.6\,$\mu$m around Cepheids as a function of the pulsation period.}
\label{correlation}
\end{figure}

\citet{Merand-2007-08} showed for the classical Cepheids a likely correlation between the pulsation period and the CSE flux (relative to the photosphere) in the $K$ band. We plotted in Fig.~\ref{correlation} the relative CSE flux versus the period at 8.6\,$\mu$m. It seems that at 8.6\,$\mu$m the longer periods have larger excesses, as concluded by \citet{Merand-2007-08} in the $K$ band. Assuming this excess is linked to a mass loss phenomena, this correlation shows that long period Cepheids have a larger mass loss than shorter period, less massive stars. Such a behavior could be explained by the stronger velocity fields in longer period Cepheids, and the presence of shock waves at certain pulsation phases \citep{Nardetto-2006-07,Nardetto-2008-10-1}.

\section{Conclusion}

We presented new thermal IR photometry and spectral energy distributions of 8 classical Cepheids and 3 type II Cepheids. We detect extended emission around 10 stars in the thermal infrared domain, of which 3 were previously known. The sample of classical Cepheids with CSEs is extended with five additional stars to a total of 9 stars. This confirms that the presence of circumstellar material around classical Cepheids is a widespread phenomenon.

A correlation is also found between the pulsation period and the CSE flux in the thermal IR and confirms the same correlation found in the $K$ band by \citet{Merand-2007-08}. It thus confirms that longer period Cepheids have apparently larger mass loss rates than shorter periods. Apart from their probable importance in the evolution of Cepheids, the existence of these CSEs may also impact Cepheid distance measurements in the infrared domain, in particular with the JWST.

We also point out the need for more optical and near-infrared Cepheid light curves. This is the main source of uncertainty of the spectral energy distributions at a given phase of pulsation. A dedicated $J, H, K$ photometric survey of few month would be sufficient for a good coverage in phase and would favour the mid- and far-infrared excess detection.


 \begin{acknowledgements}
We thank the referee for his/her suggestions that led to improvements of this article. We received the support of PHASE, the high angular resolution partnership between ONERA, Observatoire de Paris, CNRS, and University Denis Diderot Paris 7. This work made use of the SIMBAD and VIZIER astrophysical database from CDS, Strasbourg, France and the bibliographic informations from the NASA Astrophysics Data System. Data processing for this work have been done using the Yorick language which is freely available at http://yorick.sourceforge.net/. This research is based on observations with AKARI, a JAXA project with the participation of ESA.
\end{acknowledgements}


\bibliographystyle{aa}   
\bibliography{/Users/alexandregallenne/Sciences/Articles/bibliographie}


\onllongtab{2}{
\begin{longtable}{ccccccccc}
\caption{\label{log} Log of our \emph{VISIR} BURST mode observations.}\\
\hline\hline
MJD					& $\phi$	& Star								& Filter 			& DIT 	&	$N$			& seeing		&  AM	&	\#		\\
						&				&										&					& (ms)	&					& (\arcsec)	&			&			\\
\hline
\endfirsthead
\caption{continued.}\\
\hline \hline
MJD					& $\phi$	& Star								& Filter 			& DIT 	&	$N$			& seeing		&  AM	&	\#		\\
						&				&										&					& (ms)	&					& (\arcsec)	&			&			\\
\hline
\endhead
\hline
54~610.035		&				& \object{HD~89682}		&  PAH1		& 16		&	22~500		&	1.2			&	1.26	&	1		\\
54~610.042		&				&	HD~89682					&	PAH2		& 8		&	48~000		&	1.1			&	1.27	&	2		\\
54~610.056		&	0.62		&	\object{U~Car}				&	PAH1		& 16		&	22~500		&	1.5			&	1.30	&	3		\\
54~610.064		&	0.62		&	U~Car							&	PAH2		& 8		&	48~000		&	1.1			&	1.32	&	4		\\
54~610.081		&				&	\object{HD~98118}		&	PAH1		& 16		&	22~500		&	1.1			&	1.31	&	5		\\
54~610.088		&				&	HD~98118					&	PAH2		& 8		&	48~000		&	1.1			&	1.35	&	6		\\
54~610.104		&	0.65		&	\object{X~Sgr}				&	PAH1		& 16		&	22~500		&	1.0.			&	1.61	&	7		\\
54~610.111		&	0.65		&	X~Sgr							&	PAH2		& 8		&	24~000		&	1.0			&	1.53	&	8		\\
54~610.126		&	0.71		&	\object{Y~Oph}				&	PAH1		& 16		&	22~500		&	1.2			&	1.63	&	9		\\
54~610.134		&	0.71		&	Y~Oph							&	PAH2		& 8		&	48~000		&	1.8			&	1.55	&	10	\\
54~610.158		&				&	\object{HD~99998}		&	PAH1		& 16		&	22~500		&	1.0			&	1.88	&	11	\\
54~610.166		&				&	HD~99998					&	PAH2		& 8		&	48~000		&	1.2			&	2.01	&	12	\\
54~610.182		&				&	\object{HD~124294}		&	PAH1		& 16		&	22~500		&	0.9			&	1.12	&	13	\\
54~610.190		&				&	HD~124294					&	PAH2		& 8		&	48~000		&	1.0			&	1.14	&	14	\\
54~610.213		&	0.42		&	\object{W~Sgr}				&	PAH1		& 16		&	22~500		&	1.0			&	1.07	&	15	\\
54~610.220		&	0.42		&	W~Sgr							&	PAH2		& 8		&	48~000		&	1.1			&	1.06	&	16	\\
54~610.236		&	0.48		&	\object{R~Sct}				&	PAH1		& 16		&	22~500		&	1.1			& 	1.16	&	17	\\
54~610.243		&	0.48		&	R~Sct							&	PAH2		& 8		&	48~000		&	1.1			& 	1.14	&	18	\\
54~610.258		&				&	\object{HD~161096}		&	PAH1		& 16		&	22~500		&	0.9			&	1.15	&	19	\\
54~610.266		&				&	HD~161096					&	PAH2		& 8		&	48~000		&	0.8			&	1.15	&	20	\\
54~610.282		&	0.40		&	\object{$\eta$~Aql}		&	PAH1		& 16		&	22~500		&	1.0			&	1.22	&	21	\\
54~610.289		&	0.40		&	$\eta$~Aql					&	PAH2		& 8		&	48~000		&	0.8			&	1.20	&	22	\\
54~610.305		&	0.02		&	\object{SV~Vul}				&	PAH1		& 16		&	22~500		&	1.0			&	1.72	&	23	\\
54~610.312		&	0.02		&	SV~Vul							&	PAH2		& 8		&	48~000		&	1.1			&	1.69	&	24	\\
54~610.327		&				&	\object{HD~203504}		&	PAH1		& 16		&	22~500		&	0.9			&	1.69	&	25	\\
54~610.334		&				&	HD~203504					&	PAH2		& 8		&	48~000		&	0.9			&	1.63	&	26	\\
54~610.349		&	0.14		&	\object{AC~Her}			&	PAH1		& 16		&	22~500		& 	0.8			&	1.58	&	27	\\
54~610.357		&	0.14		&	AC-Her							&	PAH2		& 8		&	48~000		&	0.9			&	1.60	&	28	\\
54~610.373		&	0.71		&	Y~Oph							&	PAH1		& 16		&	22~500		&	0.8			&	1.32	&	29	\\
54~610.380		&	0.71		&	Y~Oph							&	PAH2		& 8		&	48~000		&	0.7			&	1.36	&	30	\\
54~610.395		&	0.42		&	$\eta$~Aql					&	PAH1		& 16		&	22~500		&	0.8			&	1.15	&	31	\\
54~610.402		&	0.42		&	$\eta$~Aql					&	PAH2		& 8		&	48~000		&	0.9			&	1.17	&	32	\\
54~610.417		&				&	\object{HD~196321}		&	PAH1		& 16		&	22~500		&	0.8			&	1.11	&	33	\\
54~610.424		&				&	HD~196321					&	PAH2		& 8		&	48~000		&	0.7			&	1.12	&	34	\\
54~611.016		&				&	HD~89682					&	PAH1		& 16		&	22~500		&	0.7			&	1.22	&	35	\\
54~611.024		&				&	HD~89682					&	SiC			& 20		&	18~000		&	0.8			&	1.24	&	36	\\
54~611.035		&	0.64		&	U~Car							&	PAH1		& 16		&	22~500		&	0.8			&	1.27	&	37	\\
54~611.042		&	0.64		&	U~Car							&	SiC			& 20		&	18~000		&	0.8			&	1.28	&	38	\\
54~611.054		&	0.26		&	\object{AX~Cir}				&	PAH1		& 16		&	22~500		&	0.8			&	1.40	&	39	\\
54~611.061		&	0.26		&  AX~Cir							&	SiC			& 20		&	18~000		&	0.8			&	1.38	&	40	\\
54~611.073		&				&	HD~124294					&	PAH1		& 16		&	22~500		&	0.7			&	1.06	&	41	\\
54~611.081		&				&	HD~124294					&	SiC			& 20		&	18~000		&	0.7			&	1.05	&	42	\\
54~611.093		&	0.27		&	AX~Cir							&	PAH1		& 16		&	22~500		&	0.8			&	1.32	&	43	\\
54~611.101		&	0.27		&	AX~Cir							&	SiC			& 20		&	18~000		&	0.7			&	1.31	&	44	\\
54~611.112		&	0.79		&	X~Sgr							&	PAH1		& 16		&	22~500		&	0.8			&	1.50	&	45	\\
54~611.119		&	0.79		&	X~Sgr							&	SiC			& 20		&	18~000		&	1.1			&	1.44	&	46	\\
54~611.131		&	0.54		&	W~Sgr							&	PAH1		& 16		&	22~500		&	0.8			&	1.43	&	47	\\
54~611.139		&	0.54		&	W~Sgr							&	SiC			& 20		&	18~000		&	0.8			&	1.38	&	48	\\
54~611.151		&				&	HD~161096					&	PAH1		& 16		&	22~500		&	0.8			&	1.49	&	49	\\
54~611.158		&				&	HD~161096					&	SiC			& 20		&	18~000		&	0.8			&	1.43	&	50	\\
54~611.170		&	0.76		&	Y~Oph							&	PAH1		& 16		&	22~500		&	0.7			&	1.27	&	51	\\
54~611.177		&	0.76		&	Y~Oph							&	SiC			& 20		&	18~000		&	0.7			&	1.23	&	52	\\
54~611.189		&	0.48		&	R~Sct							&	PAH1		& 16		&	22~500		&	0.6			&	1.39	&	53	\\
54~611.196		&	0.48		&	R~Sct							&	SiC			& 20		&	18~000		&	0.6			&	1.34	&	54	\\
54~611.207		&				&	HD~161096					&	PAH1		& 16		&	22~500		&	0.6			&	1.21	&	55	\\
54~611.215		&				&	HD~161096					&	SiC			& 20		&	18~000		&	0.6			&	1.19	&	56	\\
54~611.227		&	0.90		&	\object{$\kappa$~Pav}	&	PAH1		& 16		&	22~500		&	0.7			&	1.45	&	57	\\
54~611.235		&	0.90		&	$\kappa$~Pav				&	SiC			& 20		&	18~000		&	0.7			&	1.43	&	58	\\
54~611.248		&	0.53		&	$\eta$~Aql					&	PAH1		& 20		&	18~000		&	0.7			&	1.38	&	59	\\
54~611.255		&	0.53		&	$\eta$~Aql					&	SiC			& 25		&	14~400		&	0.7			&	1.33	&	60	\\
54~611.273		&				&	\object{HD~133774}		&	PAH1		& 16		&	22~500		&	0.8			&	1.35	&	61	\\
54~611.280		&				&	HD~133774					&	SiC			& 20		&	18~000		&	0.8			&	1.40	&	62	\\
54~611.310		&				&	HD~161096					&	PAH1		& 16		&	22~500		&	0.7			&	1.21	&	63	\\
54~611.317		&				&	HD~161096					&	SiC			& 20		&	18~000		&	0.7			&	1.23	&	64	\\
54~611.340		&	0.15		&	AC~Her							&	PAH1		& 16		&	22~500		&	0.8			&	1.54	&	66	\\
54~611.347		&	0.15		&	AC~Her							&	SiC			& 20		&	18~000		&	0.8			&	1.57	&	67	\\
54~611.365		&	0.04		&	SV~Vul							&	PAH1		& 16		&	22~500		&	0.7			&	1.64	&	68	\\
54~611.372		&	0.04		&	SV~Vul							&	SiC			& 20		&	18~000		&	0.8			&	1.65	&	69	\\
54~611.384		&				&	HD~203504					&	PAH1		& 16		&	22~500		&	0.7			&	1.42	&	70	\\
54~611.392		&				&	HD~203504					&	SiC			& 20		&	18~000		&	0.8			&	1.41	&	71	\\
54~611.403		&	0.62		&	\object{FF~Aql}				&	PAH1		& 16		&	22~500		&	0.6			&	1.66	&	72	\\
54~611.411		&	0.62		&	FF~Aql							&	SiC			& 20		&	18~000		&	0.7			&	1.73	&	73	\\
54~611.422		&	0.56		&	$\eta$~Aql					&	PAH1		& 16		&	22~500		&	0.7			&	1.24	&	74	\\
54~611.430		&	0.56		&	$\eta$~Aql					&	SiC			& 20		&	18~000		&	0.7			&	1.27	&	75	\\
\hline
\end{longtable}
\tablefoot{MJD is the Modified Julian Date at the start of the exposures of the target. $\phi$ is the phase of the Cepheid. DIT denotes the Detector Integration Time for one short exposure image. $N$ represents the total number of frames. The seeing is measured in the visible by the DIMM station. AM denotes the airmass.}
}

\end{document}